\documentclass[runningheads]{llncs}
\usepackage[T1]{fontenc}
\usepackage{amsmath,amssymb}
\usepackage{enumitem}
\usepackage{graphicx}
\usepackage[hidelinks]{hyperref}
\usepackage{float}
\begin{document}

\title{The Advice Complexity of Online AdWords\\ with Hard Budgets}
\titlerunning{The Advice Complexity of AdWords with Hard Budgets}
\author{Pratik Chowdhury\inst{1}}
\institute{Department of Computer Science and Engineering,\\ National Institute of Technology Durgapur, India\\
\email{pc.22cs1104@phd.nitdgp.ac.in}}
\maketitle

\begin{abstract}
Advice complexity measures how many bits of information about the future an online algorithm needs to achieve a given competitive ratio. We study the advice complexity of the Online AdWords Problem with Hard Budgets (OAW-HB) and determine several advice complexities exactly or to the first-order term. All our lower bounds rest on one method, namely an exponential moment whose base and input distribution are chosen optimally.

For small bids, BALANCE is $(1-1/e)$-competitive without advice. On the nested instances behind this bound we establish that ratio $1-\epsilon$ costs $n\log_2\frac1\epsilon-n\log_2e+o(n)$ bits as $\epsilon\to0$, for every budget size and in particular for bipartite matching, and that exact optimality costs exactly $\lceil\log_2n!\rceil$ bits. The matching algorithm does the opposite of BALANCE, i.e., it spends its whole error budget at the start to build a reserve. Beyond the nested family, on every instance whose eligible sets shrink over time, with arbitrary budgets and demands, BALANCE refined by $\lceil1/\epsilon\rceil$ priority classes attains ratio $1-\epsilon$ with $\lceil n\log_2\lceil1/\epsilon\rceil\rceil$ bits, so the advice complexity of this class is $n\log_2\frac1\epsilon\pm O(n)$. Finally, on a hidden-target family on which BALANCE is exactly $3/4$-competitive, ratio $\frac34+\epsilon$ costs $\Theta(\epsilon^2n)$ bits, even for randomized algorithms, and the optimal advice is a covering design.

\end{abstract}

\section{Introduction}
\label{sec:intro}

In the AdWords problem \cite{mehta2007adwords} queries arrive online, each query is eligible for some advertisers with given bids, and every advertiser has a hard budget. When every bid is small compared with the budget, the algorithm BALANCE is $(1-1/e)$-competitive, and this is optimal even for randomized algorithms \cite{mehta2007adwords}. The loss comes entirely from not knowing the future. The \emph{advice model} \cite{dobrev2008much,hromkovivc2010information,bockenhauer2009advice,boyar2017online} quantifies it. An oracle that sees the whole input writes a bit string, and the \emph{advice complexity} of ratio $\rho$ is the least number of bits that an online algorithm must read to be $\rho$-competitive.

Advice lower bounds are usually proved by counting or by a reduction from string guessing \cite{bockenhauer2014string,boyar2015advice}, and they are usually stated up to constant factors. For bipartite matching, for instance, ratio $1-\epsilon$ is known to need $\Omega(n\log\frac1\epsilon)$ bits, and $O(n/\epsilon^5)$ bits suffice \cite{durr2016power}. Exact constants are rare, and a notable exception is asymmetric string guessing, whose advice complexity Boyar et al.\ \cite{boyar2015advice} determined exactly. Motivated by this, we study the advice complexity of budgeted allocation and determine it exactly, to the second-order term, or to the first-order term on the instance families that underlie the known lower bounds.

\subsection{The Online AdWords Problem with Hard Budgets}

An instance $I=(U,V,E,B,b)$ of OAW-HB consists of a set $U=\{1,\dots,n\}$ of advertisers, advertiser $i$ having an integer budget $B_i$, a sequence $V$ of $m$ queries that arrive one by one, and bids $b_{ij}\ge0$ on the edges of a bipartite graph $G=(U,V,E)$. When query $j$ arrives, its neighborhood $N(j)$ and its bids are revealed. The algorithm then assigns $j$ irrevocably to some $i\in N(j)$ whose spent amount plus $b_{ij}$ does not exceed $B_i$, collecting $b_{ij}$, or it rejects $j$. We write $\mathrm{ALG}(I)$ and $\mathrm{OPT}(I)$ for the revenue of the online algorithm and of an optimal offline allocation. With unit bids and unit budgets the problem is online bipartite matching. Under the small-bid assumption every bid is small compared with the budget, and in the limit the demand becomes divisible.

\subsection{Our Contributions}

Our results are summarized in Table~\ref{tab:results}.

\paragraph{1. The exact cost of ratio $1-\epsilon$ for small bids.} Let $N_\pi(n,B)$ be the nested instance behind the bound $1-1/e$. It has $n$ advertisers of budget $B$ and $n$ phases of $B$ unit-bid queries, and phase $t$ is eligible for the advertisers $\pi(t),\dots,\pi(n)$ of a hidden order $\pi$. We establish (Theorems~\ref{thm:nested-lb} and~\ref{thm:nested-ub}, Corollary~\ref{cor:nested-sharp}) that ratio $1-\epsilon$ on this family costs
\[
b^*(n,\epsilon)\ =\ n\log_2\frac1\epsilon-n\log_2e\ \pm\ O\Big(\epsilon n\log\frac1\epsilon+\frac1\epsilon+\log n\Big)
\]
bits, for every budget $B\ge1$ (so in particular for bipartite matching) and in the divisible model as well. Exact optimality costs exactly $\lceil\log_2n!\rceil$ bits for every $B$ (Corollary~\ref{cor:nested-exact}), which extends the bound of Miyazaki for matching \cite{miyazaki2014advice}. The optimal advice is not what one might expect. Instead of spreading the demand as BALANCE does, the optimal algorithm spends its whole error budget at the start to build a reserve of full advertisers, after which the advice only has to guarantee that each exit falls into a \emph{sliding} window of the size of the reserve (Figure~\ref{fig:buffer}).

\begin{figure}[!ht]
\centering
\includegraphics[width=0.78\textwidth,height=0.22\textheight,keepaspectratio]{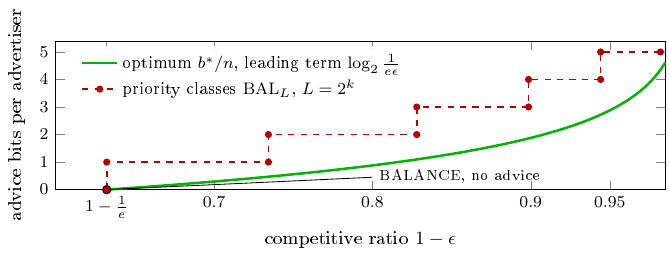}
\caption{\footnotesize The price of competitiveness on the nested family $N_\pi(n,B)$. The solid curve is the leading term $\log_2\frac1{e\epsilon}$ of $b^*(n,\epsilon)/n$, sharp as $\epsilon\to0$ (Corollary~\ref{cor:nested-sharp}). The dashed steps show $\log_2L$ bits for $L=2^k$ priority classes, plotted at their exact asymptotic ratios $1-r_{L+1}/L$ (Proposition~\ref{prop:pc-exact}). At the corners, the gap tends to $\log_2e$ bits (Remark~\ref{rem:windows}). The leading term vanishes at $\epsilon=1/e$, the ratio of BALANCE; we do not claim sharpness near $1-1/e$.}
\label{fig:price}
\end{figure}

\begin{figure}[!ht]
\centering
\includegraphics{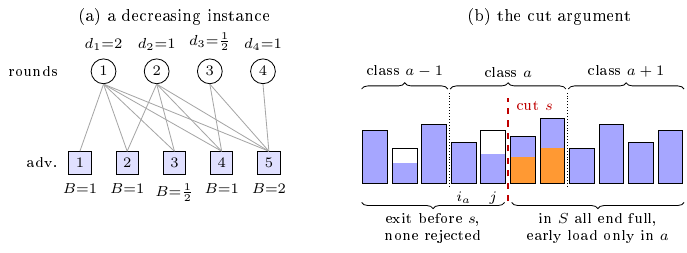}
\caption{(a) A decreasing instance as a bipartite graph. Round $t$ brings demand $d_t$ that is eligible for every advertiser exiting at or after $t$ (here $\tau=(1,2,2,3,4)$), so the eligible sets shrink. (b) The quantities in the proof of Theorem~\ref{thm:pc-all}. Advertisers are drawn in the exit order $\prec$, a box shows the budget and the shading the final load, and dotted lines separate the classes. Advertiser $j$ is the last one that ends with unused budget, and the cut is at $s=\tau_j+1$. Everything after the cut ends full, nothing is rejected before it, and demand placed on $S$ before round $s$ (orange) can only go to the class of $j$, never to its first member $i_a$. So the loss relative to an optimal solution $x$ is at most $x(a^-)$.}
\label{fig:cut}
\end{figure}

\paragraph{2. Every decreasing instance.} An instance is \emph{decreasing} if its eligible sets shrink along the stream, as when campaigns end one after another, while budgets, exit times and demands are arbitrary. The natural advice is a priority class for each advertiser, which is used by BALANCE refined into $L$ priority classes. We prove that on every decreasing instance this algorithm loses at most the optimal load of one class, minus one advertiser (Theorem~\ref{thm:pc-all}). The proof is a single cut argument (Figure~\ref{fig:cut}). With classes cut along an optimal solution, $\lceil n\log_2\lceil1/\epsilon\rceil\rceil$ bits give ratio $1-\epsilon$ (Corollary~\ref{cor:dec}). Since the nested family is decreasing, the advice complexity of decreasing instances is $n\log_2\frac1\epsilon\pm O(n)$, and exact optimality on them costs exactly $\lceil\log_2n!\rceil$ bits. On the nested family, priority classes exceed the optimum $b^*$ by at most $\log_2e+\log_2(1+\epsilon)$ bits per advertiser, up to the error terms of Corollary~\ref{cor:nested-sharp}. These $\log_2e$ bits are exactly the price of \emph{fixed} instead of \emph{sliding} windows (Remark~\ref{rem:windows} and Figure~\ref{fig:price}).

\vspace{-1cm}

\vspace{-1cm}
\paragraph{3. Hidden targets above $3/4$.} We also consider a two-phase family in which flexible demand must be placed before it is revealed which advertisers will receive rigid demand. On this family BALANCE is exactly $3/4$-competitive. We prove that ratio $\frac34+\epsilon$ needs, and can be achieved with, $\Theta(\epsilon^2n)$ bits, for every budget and even against randomized algorithms (Theorems~\ref{thm:hidden} and~\ref{thm:hidden-upper}, Corollary~\ref{cor:hidden}). Moreover the advice complexity of this family equals the logarithm of a covering number with a defect (Theorem~\ref{thm:char}).

\begin{table}[H]
\caption{Main results, with $n$ advertisers and $m$ queries. The $o(n)$ term in the first row holds as $\epsilon\to0$ and $\epsilon n\to\infty$, and the bound $\Omega(n\log\frac mn)$ holds when $m/n$ is large enough.}
\label{tab:results}
\centering
\scriptsize
\setlength{\tabcolsep}{2pt}
\renewcommand{\arraystretch}{0.9}

\resizebox{0.95\textwidth}{!}{%
\begin{tabular}{@{}p{2.5cm}p{1.3cm}p{3.3cm}l@{}}
\hline
\textbf{Setting} & \textbf{Ratio} & \textbf{Advice (bits)} & \textbf{Ref.}\\
\hline
Small bids, nested & $1-\epsilon$ & $n\log_2\frac1{e\epsilon}\pm o(n)$ & Cor.~\ref{cor:nested-sharp}\\
 & exact & $\lceil\log_2n!\rceil$ & Cor.~\ref{cor:nested-exact}\\
Small bids, decreasing & $1-\epsilon$ & $n\log_2\frac1\epsilon\pm O(n)$ & Cor.~\ref{cor:dec}\\
Small bids & $1-\frac1e+\delta$ & $\Omega(\delta^3n)$ & Thm.~\ref{thm:balance-barrier}\\
Hidden targets & $\frac34+\epsilon$ & $\Theta(\epsilon^2n)$, randomized & Cor.~\ref{cor:hidden}\\
Large bids & $\frac16$ and $\rho>\frac12$ & $1$ and $\Omega(n\log\frac mn)$ & Thm.~\ref{thm:one-bit},~\ref{thm:half-barrier}\\
\hline
\end{tabular}%
}
\end{table}

\subsection{Techniques}

All our lower bounds rest on a classical principle \cite{emek2011online,mikkelsen2016randomization}. An algorithm that reads $b$ bits is one of $2^b$ advice-free algorithms. So if a random input defeats every advice-free algorithm except with probability $P$, then $b\ge\log_2(1/P)$. We bound $P$ \emph{optimally}, by an exponential moment whose base and input distribution make the bound for each round tight.

For AdWords the payoff of an exit depends on everything the algorithm did before, and the moment becomes computable through the \emph{reserve law}. The budget filled on advertisers that are still present is at most one phase of demand plus the loss incurred so far, so an algorithm can hedge only to the extent that it has already lost. The same law suggests the optimal algorithm, which loses early on purpose. For priority classes we look at the last advertiser that ends with unused budget. Every later advertiser ends full, nothing is rejected before it leaves, and every unit placed early on a later advertiser stays inside its class, so the cut at its exit bounds the loss by one class. For hidden targets the upper bound is a covering family built from an anti-concentration bound for the hypergeometric distribution.

\subsection{Related Works}

Advice complexity was introduced in \cite{dobrev2008much,hromkovivc2010information,bockenhauer2009advice}, and we refer to the survey \cite{boyar2017online}. String guessing \cite{bockenhauer2014string} and its asymmetric version $\mathrm{ASG}$ \cite{boyar2015advice} are the standard sources of lower bounds. Boyar et al.\ \cite{boyar2015advice} determined the advice complexity of $\mathrm{ASG}$ via covering designs and defined the class $\mathrm{AOC}$. For online bipartite matching, Miyazaki \cite{miyazaki2014advice} proved that $\lceil\log_2n!\rceil$ bits are necessary and sufficient for optimality. D\"urr et al.\ \cite{durr2016power} showed that ratio $1-\epsilon$ needs $\Omega(n\log\frac1\epsilon)$ bits and admits $O(n/\epsilon^5)$ bits, and beating ratio $1/2$ costs $\Theta(\log\log n)$ bits \cite{pena2016limitations,buchbinder2023lossless}. Online knapsack is AdWords with one advertiser and private queries, and for it one bit gives ratio $1/2$ while beating $1/2$ costs about $\log m$ bits \cite{bockenhauer2012advice}. To the best of our knowledge, the advice complexity of AdWords has not been studied before.

\section{Preliminaries}
\label{sec:prelim}

An oracle that knows the instance writes an infinite tape, and a deterministic online algorithm reads a prefix of it. An algorithm that reads at most $b$ bits behaves on every instance as one of the $2^b$ advice-free algorithms $\mathcal A_\sigma$ that read the tape $\sigma0^\infty$, $\sigma\in\{0,1\}^b$. Its decisions on a common prefix of two instances coincide whenever it reads the same advice, and we call this \emph{execution invariance}. Two generic tools are used throughout.

\begin{lemma}
\label{lem:tilt}
(i) Let $\mu$ be a distribution on a family $\mathcal F$ of instances such that every advice-free deterministic algorithm meets a given guarantee on a $\mu$-random instance with probability at most $P$. Then every deterministic algorithm that meets the guarantee on all of $\mathcal F$ reads at least $\log_2(1/P)$ bits on some instance.
(ii) Let $Z_1,\dots,Z_n$ take values in $\mathbb R\cup\{+\infty\}$ and be adapted to a filtration $(\mathcal G_i)$, and let $u\in(0,1)$ and $c_i\ge0$ satisfy $\mathbb E[u^{Z_i}\mid\mathcal G_{i-1}]\le c_i$ (with $u^{+\infty}=0$). Then $\Pr[\sum_iZ_i\le z]\le u^{-z}\prod_ic_i$ for all real $z$.
\end{lemma}
\begin{proof}
(i) The success events of the $2^b$ algorithms $\mathcal A_\sigma$ cover $\mathcal F$, so $1\le2^bP$. (ii) Since $u<1$, the event $\sum_iZ_i\le z$ means $u^{\sum_iZ_i}\ge u^z$. By the tower property $\mathbb E[u^{\sum_iZ_i}]\le\prod_ic_i$, and Markov's inequality gives the claim. \qed
\end{proof}

\begin{lemma}
\label{lem:cover}
Let $\mathcal F$ be a finite family of instances and $\mathcal D$ a distribution over deterministic advice-free algorithms such that every $I\in\mathcal F$ is served by a $\mathcal D$-random algorithm with probability at least $q>0$. Then some $N\le\ln|\mathcal F|/q+1$ algorithms serve all of $\mathcal F$ between them, so $\lceil\log_2N\rceil$ bits of advice suffice.
\end{lemma}
\begin{proof}
For $N$ independent draws the expected number of unserved instances is at most $|\mathcal F|(1-q)^N\le|\mathcal F|e^{-qN}<1$ once $N>\ln|\mathcal F|/q$. The oracle names a serving algorithm. \qed
\end{proof}

\section{The Exact Cost of Ratio $1-\epsilon$ for Small Bids}
\label{sec:nested}

Until Section~\ref{sec:nested-pc} the instance is $N_\pi(n,B)$ from Section~\ref{sec:intro}. Advertiser $\pi(t)$ \emph{exits} after phase $t$, $\mathrm{OPT}=nB$, and ratio $1-\epsilon$ means losing at most $\epsilon nB$ queries. In the divisible model a phase is a parcel of mass $B$ that may be split arbitrarily, and indivisible algorithms are a special case.

\subsection{The Reserve Law and the Lower Bound}

\begin{theorem}
\label{thm:nested-lb}
Let $n\ge1$, $B\ge1$, $\epsilon\in(0,\frac14]$ and $K=\epsilon n$. Every deterministic algorithm that is $(1-\epsilon)$-competitive on all instances $N_\pi(n,B)$, divisible or not, reads at least
\[
\log_2\frac1{P_n(K)}\ \ge\ n\log_2\frac n{e(K+1)}-2K\log_2\frac{n+1}K-0.45\,K
\]
bits, where $P_n(K)=\inf_{s>0}e^{-s(n-K)}\prod_{\ell=1}^n\big(1+(e^s-1)\min\{1,\frac{K+1}\ell\}\big)$.
\end{theorem}
\begin{proof}
Normalize the budgets to $1$, fix an advice-free algorithm, and let $\pi$ be uniform. Let $X_t\in[0,1]$ be the fill of $\pi(t)$ when it exits. It receives nothing later, so the revenue is $\sum_tX_t$. Put $D_t=\sum_{s\le t}(1-X_s)$, so that success means $D_n\le K$.

(a) By execution invariance, the loads after phase $t$ depend only on the exits $\pi(1),\dots,\pi(t-1)$. Given these, $\pi(t)$ is uniform among the $\ell_t=n-t+1$ live advertisers.

(b) \emph{Reserve law.} After phase $t$ at most $t$ units have been placed, and $\sum_{s<t}X_s$ of them sit on exited advertisers. So the live advertisers hold at most $1+D_{t-1}$.

Hence $\mathbb E[X_t\mid\pi_{<t}]\le\min\{1,(1+D_{t-1})/\ell_t\}$. Let $Y_t=X_t\mathbf 1[D_{t-1}\le K]$. Success forces $\sum_tY_t\ge n-K$, and by convexity of $y\mapsto e^{sy}$ on $[0,1]$,
\[
\mathbb E\big[e^{sY_t}\mid\pi_{<t}\big]\le1+(e^s-1)\min\Big\{1,\frac{K+1}{\ell_t}\Big\}.
\]
Lemma~\ref{lem:tilt} with $u=e^{-s}$ and $Z_t=-Y_t$ gives the first bound.

For the explicit form let $M=K+1$ and $m=\lfloor M\rfloor$. The factors with $\ell\le m$ equal $e^s$. For $\ell>m$ we use $\ln(1+(e^s-1)a)\le s+\ln a+e^{-s}/a$ with $a=M/\ell$. This gives $\ln P_n(K)\le sK+\sum_{\ell>m}\ln\frac M\ell+e^{-s}\frac{n(n+1)}{2M}$. The sum is at most $-n\ln\frac n{eM}$, by $\ln n!\ge n\ln\frac ne$ and $m!\le M^m$. The choice $e^{-s}=\frac{2MK}{n(n+1)}$ bounds the other two terms by $2K\ln\frac{n+1}K+(1-\ln2)K$. \qed
\end{proof}

\subsection{The Buffer Algorithm, or Losing Early on Purpose}

\begin{figure}[!ht]
\centering
\includegraphics[width=\linewidth]{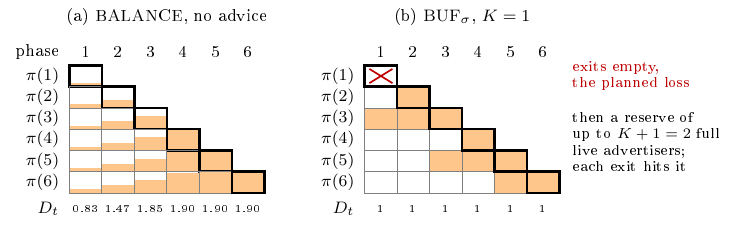}
\caption{The two strategies on $N_\pi(6,1)$. Row $\pi(i)$ is the advertiser that exits after phase $i$. A cell shows its fill after the phase has been placed, the thick cell is its exit, and its unused budget there is lost. $D_t$ is the loss after $t$ exits. (a) BALANCE spreads every phase over all live advertisers, so every early exit wastes budget, and it loses $1.9$. (b) $\mathrm{BUF}_\sigma$ with $\sigma=(\pi(3),\pi(2),\pi(5),\pi(4),\pi(6),\pi(1))$, on an order in the event $E_\sigma(\pi)$ of Theorem~\ref{thm:nested-ub}. It spends its whole error budget $K=1$ on the first exit and then keeps a sliding reserve of $\min\{K+1,n-t+1\}$ full advertisers, so it loses $1$. Until the first rejection (phase $5$ in (a) and phase $6$ in (b)) the live load after phase $t$ equals $1+D_{t-1}$, the most that the reserve law allows, and the two strategies differ only in how they distribute it.}
\label{fig:buffer}
\end{figure}

For a priority order $\sigma$, algorithm $\mathrm{BUF}_\sigma$ assigns every query to the first advertiser in the order $\sigma$ that is eligible and has room, and it rejects the query if there is none. Since a phase brings exactly one budget of demand, every advertiser is either empty or full at the end of every phase. The algorithm does not use $B$, and for $B=1$ it is a matching algorithm. Figure~\ref{fig:buffer} compares it with BALANCE.

\begin{theorem}
\label{thm:nested-ub}
Let $n\ge2$, $B\ge1$ and let $K$ be an integer with $1\le K\le n/4$. There is a deterministic algorithm that loses at most $KB$ queries on every instance $N_\pi(n,B)$ and reads at most
\[
n\log_2\frac n{e(K+1)}-K\log_2\frac n{K+1}+\Big(2K+\frac{K(K+1)}n+2\Big)\log_2e+\tfrac52\log_2n+1
\]
bits of advice.
\end{theorem}
\begin{proof}
Let $F_t$ be the set of live full advertisers after phase $t$ has been placed, and let $E_\sigma(\pi)$ be the event that $\pi(t)\notin F_t$ for $t\le K$ and $\pi(t)\in F_t$ for $t>K$. On $E_\sigma(\pi)$ the following holds by induction.
\begin{itemize}
\item For $t\le K$, $F_t$ consists of the first $t$ advertisers in the order $\sigma$ among those that have not exited before phase $t$, and the exiting advertiser is empty.
\item For $t>K$, $|F_t|=\min\{K+1,n-t+1\}$, since a full advertiser exits and the next phase fills the next empty one.
\end{itemize}
So exactly $K$ advertisers exit empty, and the loss is $KB$. By (a) in the proof of Theorem~\ref{thm:nested-lb}, for uniform $\pi$
\[
\Pr[E_\sigma(\pi)]=\prod_{t=1}^K\Big(1-\frac t{n-t+1}\Big)\prod_{\ell=K+2}^{n-K}\frac{K+1}\ell=:P.
\]
The event depends only on the relative order of $\sigma$ and $\pi$, so for fixed $\pi$ and uniform $\sigma$ it also has probability $P$. Lemma~\ref{lem:cover} over the $n!$ instances gives advice $\log_2(1/P)+\log_2(\ln n!+1)+1\le\log_2(1/P)+2\log_2n+1$.

The estimate of $\log_2\frac1P$ that gives the stated bound is in Appendix~\ref{app:nested}. \qed
\end{proof}

\begin{corollary}
\label{cor:nested-sharp}
For all $B\ge1$, $\epsilon\in(0,\frac14]$ and $n\ge1/\epsilon$,
\[
\Big|\,b^*(n,\epsilon)-n\log_2\frac1{e\epsilon}\,\Big|\ \le\ 3\epsilon n\log_2\frac2\epsilon+\frac2\epsilon+3\log_2n+5,
\]
where $b^*(n,\epsilon)$ is the least advice for ratio $1-\epsilon$ on $N_\pi(n,B)$. Hence $b^*(n,\epsilon)=n\log_2\frac1\epsilon-n\log_2e+o(n)$ as $\epsilon\to0$ and $\epsilon n\to\infty$.
\end{corollary}
\begin{proof}
For the lower bound we take $K=\epsilon n$ in Theorem~\ref{thm:nested-lb} and use three estimates, namely $n\log_2\frac n{e(K+1)}\ge n\log_2\frac1{e\epsilon}-\frac{\log_2e}\epsilon$, $2K\log_2\frac{n+1}K\le2\epsilon n\log_2\frac2\epsilon$, and $0.45K\le0.15\epsilon n\log_2\frac2\epsilon$.

For the upper bound we take $K=\lfloor\epsilon n\rfloor$ in Theorem~\ref{thm:nested-ub}. There $K+1>\epsilon n$ bounds the leading term by $n\log_2\frac1{e\epsilon}$, the second term is non-positive, and $(2K+\frac{K(K+1)}n+2)\log_2e\le3.25\epsilon n+3.25$. \qed
\end{proof}

\begin{corollary}
\label{cor:nested-exact}
For all $n,B\ge1$, exact optimality on $N_\pi(n,B)$ costs exactly $\lceil\log_2n!\rceil$ bits, divisible or not.
\end{corollary}
\begin{proof}
With $K=0$ the bound in the proof of Theorem~\ref{thm:nested-lb} is $\prod_\ell(e^{-s}+(1-e^{-s})/\ell)\to1/n!$ as $s\to\infty$. Writing $\pi$ suffices for the upper bound. \qed
\end{proof}

\subsection{Priority Classes on Every Decreasing Instance}
\label{sec:nested-pc}

An instance is \emph{decreasing} if its demand arrives in rounds $1,\dots,T$, where round $t$ brings demand $d_t\ge0$ that is eligible for a set $E_t$ of advertisers, with $E_1\supseteq\dots\supseteq E_T$. Budgets $B_i>0$ are arbitrary. As above we work in the divisible model, and a unit of demand earns $1$ wherever it is placed. The nested family is the case $d_t=B$ and $E_t=\{\pi(t),\dots,\pi(n)\}$. Advertiser $i$ \emph{exits} at $\tau_i=\max\{t:i\in E_t\}$, with $\tau_i=0$ if $i\notin E_1$. We fix a total order $\prec$ of the advertisers such that $\tau_i<\tau_j$ implies $i\prec j$. A \emph{monotone labeling} gives every advertiser a class $c(i)\in\{1,\dots,L\}$ such that $i\prec j$ implies $c(i)\le c(j)$. For a class $a$ let $a^-$ be $a$ without its $\prec$-first member. A \emph{solution} $x$ assigns the demand offline, $x_i$ being the amount that advertiser $i$ receives, and $x(S)=\sum_{i\in S}x_i$. The advice gives every advertiser its class, which costs $\lceil n\log_2L\rceil$ bits. Algorithm $\mathrm{BAL}_L$ routes demand to an eligible advertiser that is not full and minimizes (class, fill fraction) lexicographically. For $L=1$ this is BALANCE.

\begin{theorem}
\label{thm:pc-all}
On every decreasing instance, for every monotone labeling and every optimal solution $x$, $\mathrm{BAL}_L$ serves at least $\mathrm{OPT}-\max_ax(a^-)$. This holds for every algorithm that routes demand to an eligible non-full advertiser of the smallest class and rejects demand only when every eligible advertiser is full, whatever it does within a class.
\end{theorem}
\begin{proof}
Let $f_i$ be the final load of advertiser $i$, so that the algorithm serves $\sum_if_i$. If every advertiser ends full, it serves $\sum_iB_i\ge\mathrm{OPT}$. Otherwise let $j$ be the $\prec$-last advertiser that ends with unused budget, let $a=c(j)$ and $s=\tau_j+1$, and let $S=\{i:\tau_i\ge s\}$ be the set of advertisers eligible from round $s$ on. Then the following hold.
\begin{itemize}
\item[(a)] Every $i\in S$ satisfies $j\prec i$, so it ends full.
\item[(b)] Nothing is rejected before round $s$, since $j$ is eligible in every round $t\le\tau_j$ and is never full.
\item[(c)] If demand is placed on some $i\in S$ in a round $t<s$, then $c(i)\le c(j)$, because $j$ was an eligible non-full advertiser at that moment, and $c(i)\ge c(j)$, because $j\prec i$. So $i\in S\cap a$.
\end{itemize}
Let $A=\sum_{t<s}d_t$, and let $p$ be the demand placed on $S$ before round $s$. Advertisers outside $S$ receive nothing from round $s$ on, so by (b) they end with $A-p$ in total. By (a) the advertisers in $S$ end with $\sum_{i\in S}B_i$, and by (c) $p\le\sum_{i\in S\cap a}B_i$. Moreover $S\cap a\subseteq a^-$, since the first member of $a$ precedes or equals $j$ and so exits before $s$. Hence the algorithm serves
\[
A-p+\sum_{i\in S}B_i\ \ge\ A+\sum_{i\in S\setminus a^-}B_i .
\]
On the other hand, $x$ restricted to the advertisers outside $a^-$ is a solution that serves $\mathrm{OPT}-x(a^-)$. It serves at most $A$ before round $s$ and at most $\sum_{i\in S\setminus a^-}B_i$ from round $s$ on, so $\mathrm{OPT}-x(a^-)\le A+\sum_{i\in S\setminus a^-}B_i$, which is at most what the algorithm serves. \qed
\end{proof}

With singleton classes the bound is $\mathrm{OPT}$, i.e., serving the earliest exit first is optimal. The proof also applies verbatim to unit bids and integral budgets (online $b$-matching), where every advertiser that is not full can accept a query.

\begin{corollary}
\label{cor:dec}
Let $b_{\mathrm{dec}}(n,\epsilon)$ be the least advice with which a deterministic algorithm has ratio at least $1-\epsilon$ on all decreasing instances with $n$ advertisers. Then $b_{\mathrm{dec}}(n,0)=\lceil\log_2n!\rceil$ and, for $0<\epsilon<1$,
\[
b_{\mathrm{dec}}(n,\epsilon)\ \le\ \Big\lceil n\log_2\Big\lceil\frac1\epsilon\Big\rceil\Big\rceil\ \le\ n\log_2\frac1\epsilon+n\log_2(1+\epsilon)+1 .
\]
For $\epsilon\le\frac14$ and $n\ge1/\epsilon$, $b_{\mathrm{dec}}(n,\epsilon)\ge n\log_2\frac1{e\epsilon}-3\epsilon n\log_2\frac2\epsilon-\frac2\epsilon-3\log_2n-5$. Hence $b_{\mathrm{dec}}(n,\epsilon)=n\log_2\frac1\epsilon\pm O(n)$.
\end{corollary}
\begin{proof}
For the upper bound we fix an optimal solution $x$ and go through the advertisers in the order $\prec$. We open a new class whenever adding the current advertiser to the current class would make $x(a^-)$ exceed $\epsilon\,\mathrm{OPT}$. Every class then has $x(a^-)\le\epsilon\,\mathrm{OPT}$, so Theorem~\ref{thm:pc-all} gives ratio $1-\epsilon$. For consecutive classes $a$ and $a+1$ we have $x(a^-)+x_{i_{a+1}}>\epsilon\,\mathrm{OPT}$, where $i_{a+1}$ opens class $a+1$. These sets are disjoint for different $a$, so $(L-1)\epsilon\,\mathrm{OPT}<\mathrm{OPT}$ and $L\le\lceil1/\epsilon\rceil$. The oracle writes the labels as one number in base $\lceil1/\epsilon\rceil$. For $\epsilon=0$ singleton classes give $\mathrm{OPT}$, and the oracle writes $\prec$ in $\lceil\log_2n!\rceil$ bits. The lower bounds hold because the nested family is decreasing (Corollaries~\ref{cor:nested-sharp} and~\ref{cor:nested-exact}). \qed
\end{proof}

On the nested family, the corollary exceeds the optimum $b^*(n,\epsilon)$ by at most $n(\log_2e+\log_2(1+\epsilon))$ bits plus the error terms of Corollary~\ref{cor:nested-sharp}. There the loss of $\mathrm{BAL}_L$ telescopes over the classes and can be computed exactly. The ratio tends to $1-r_{L+1}/L$ with $r_1=0$ and $r_{a+1}=e^{r_a-1}$, which is $1-1/e$ for $L=1$ (Appendix~\ref{app:priority} and Figure~\ref{fig:price}). A demand-based variant of Theorem~\ref{thm:pc-all} also holds for indivisible bids of size at most $\gamma B_i$, provided the demand of every window is at most $(1-\gamma)$ times the capacity of the next class (Appendix~\ref{app:priority}).

\begin{remark}
\label{rem:windows}
Priority classes tell each advertiser in which of $L$ \emph{fixed} windows it exits, and about $((n/L)!)^L\approx(\frac n{eL})^n$ orders are consistent with such advice. The buffer advice only requires each exit to fall into a \emph{sliding} window of $K+1$ reserved advertisers, which about $(K+1)^n$ orders satisfy. With $K\approx n/L$ the ratio is $e^n$, which accounts exactly for the lost $\log_2e$ bits per advertiser. Whether a sliding-window rule works on every decreasing instance, as priority classes do, is open.
\end{remark}

\section{Hidden Targets and the Exact Margin above $3/4$}
\label{sec:hidden}

Let $N=2k$ advertisers have budget $B$, and let $\mathcal S$ be the family of $k$-subsets of $[N]$. For $S\in\mathcal S$ the instance $X(S)$ has two phases.
\begin{itemize}
\item \emph{Phase 1 (flexible demand).} $kB$ queries arrive, each with bid $1$ for every advertiser.
\item \emph{Phase 2 (hidden targets).} In variant $\mathsf U$ every $v\in S$ receives $B$ private queries of bid $1$, and in variant $\mathsf H$ every $v\in S$ receives one private query of bid $B$.
\end{itemize}
Filling $[N]\setminus S$ in Phase 1 serves everything, so $\mathrm{OPT}=2kB$. For even $B$, BALANCE leaves every advertiser half full after Phase 1. It then earns ratio exactly $3/4$ in variant $\mathsf U$ and $1/2$ in variant $\mathsf H$, for every $S$. The question is how much information about $S$ improves on this.

For an advice-free algorithm let $y\in\{0,\dots,B\}^N$ be its Phase-1 spending, which does not depend on $S$, and let $s=\sum_vy_v\le kB$. Let $T=\{v:y_v=0\}$ and $t=N-|T|$. In variant $\mathsf U$ a target $v$ earns at most $B-y_v$, and in variant $\mathsf H$ it earns only if $y_v=0$. So the ratio on $X(S)$ is at most
\begin{equation}\label{eq:rho}
\rho_{\mathsf U}(S)=\frac12+\frac{s-\sum_{v\in S}y_v}{2kB},\qquad \rho_{\mathsf H}(S)=\frac s{2kB}+\frac{|S\cap T|}{2k}.
\end{equation}

\begin{theorem}
\label{thm:hidden}
Let $\epsilon\in(0,\frac14)$, $B\ge1$, and let the variant be $\mathsf U$ or $\mathsf H$. Suppose that
\[
b\ \le\ \frac{\epsilon^2}{\ln2}\,N-\log_2\frac2\epsilon .
\]
Then for every randomized algorithm that reads at most $b$ bits of advice there is a target set $S\in\mathcal S$ on which its expected ratio is at most $\frac34+\frac58\epsilon<\frac34+\epsilon$. Here a randomized algorithm with advice is a probability distribution over deterministic algorithms with advice, and the advice may depend on the instance and even on the random choice.
\end{theorem}
\begin{proof}

\emph{Step 1.} Fix a deterministic advice-free algorithm, with Phase-1 spending $y$ as above, and let $S$ be uniform on $\mathcal S$. We show that
\begin{equation}\label{eq:hidden-tail}
\Pr_S\Big[\rho(S)>\tfrac34+\tfrac\epsilon2\Big]\ \le\ e^{-\epsilon^2N},
\end{equation}
where $\rho$ is $\rho_{\mathsf U}$ or $\rho_{\mathsf H}$ from \eqref{eq:rho}.
\begin{itemize}
\item \emph{Variant $\mathsf U$.} Let $Z=\sum_{v\in S}y_v$. Then $\rho_{\mathsf U}(S)>\frac34+\frac\epsilon2$ means $s-Z>(\frac12+\epsilon)kB$, and since $s\le kB$ this implies $Z<\frac s2-\epsilon kB$. Each advertiser lies in $S$ with probability $\frac12$, so $\mathbb EZ=\frac s2$. Moreover $Z$ is a sum of $k$ values drawn without replacement from the population $\{y_v\}\subseteq[0,B]$. Hoeffding's inequality for sampling without replacement \cite[Thm.~4]{hoeffding1963} gives $\Pr[Z\le\mathbb EZ-\epsilon kB]\le\exp(-2(\epsilon kB)^2/(kB^2))=e^{-2\epsilon^2k}=e^{-\epsilon^2N}$.
\item \emph{Variant $\mathsf H$.} Let $Q=|S\cap T|$, so $\mathbb EQ=\frac{|T|}2=k-\frac t2$. Only the $t$ touched advertisers spend, each at most $B$, and the Phase-1 demand is $kB$, so $s/B\le\min\{k,t\}\le\frac{k+t}2$. Then $\rho_{\mathsf H}(S)>\frac34+\frac\epsilon2$ means $Q>(\frac32+\epsilon)k-\frac sB$, which implies $Q>k-\frac t2+\epsilon k=\mathbb EQ+\epsilon k$. Hoeffding's inequality with values in $\{0,1\}$ gives the same bound $e^{-\epsilon^2N}$.
\end{itemize}

\emph{Step 2.} A deterministic algorithm that reads at most $b$ bits behaves, on every instance, like one of the $2^b$ advice-free algorithms $\mathcal A_\sigma$ obtained by fixing its first $b$ advice bits $\sigma$ (Section~\ref{sec:prelim}). Whatever advice it receives on $X(S)$, its ratio there is at most $\max_\sigma\rho_\sigma(S)$, where $\rho_\sigma$ is \eqref{eq:rho} for $\mathcal A_\sigma$. Let $G$ be the event that $\rho_\sigma(S)\le\frac34+\frac\epsilon2$ for every $\sigma$. By \eqref{eq:hidden-tail}, a union bound and the hypothesis on $b$,
\[
\Pr[\overline G]\ \le\ 2^be^{-\epsilon^2N}\ \le\ \frac\epsilon2 .
\]
Every ratio is at most $1$, so
\[
\mathbb E_S[\text{ratio}]\ \le\ \Big(\frac34+\frac\epsilon2\Big)+\Pr[\overline G]\Big(1-\frac34-\frac\epsilon2\Big)\ \le\ \frac34+\frac\epsilon2+\frac\epsilon2\Big(\frac14-\frac\epsilon2\Big)\ \le\ \frac34+\frac58\epsilon .
\]

\emph{Step 3.} A randomized algorithm is a distribution over deterministic algorithms, each reading at most $b$ bits, and Step~2 applies to each of them whatever its advice function. Averaging over the random choice gives $\mathbb E_S\mathbb E_r[\text{ratio}]\le\frac34+\frac58\epsilon$. Hence some $S$ has $\mathbb E_r[\text{ratio}]\le\frac34+\frac58\epsilon$. \qed
\end{proof}

\begin{theorem}
\label{thm:hidden-upper}
Let $k$ be even, $B\ge1$, and let $t$ be an integer with $0\le t\le k/4$. Put
\[
M=\Big\lceil(2k+1)(k+1)^2\ln2\cdot e^{16t^2/(3k)}\Big\rceil .
\]
There is a deterministic algorithm that reads $\lceil\log_2M\rceil$ bits of advice and has ratio at least $\frac34+\frac tN$ on every instance of either variant.
\end{theorem}
\noindent The advice names a \emph{reservation set} $R$ of size $k$ that meets $S$ in at most $\frac k2-t$ elements, and the algorithm fills $R$ in Phase~1. The proof is in Appendix~\ref{app:hidden}.

\begin{corollary}
\label{cor:hidden}
Let $k\ge8$ be even, $N=2k$, $B\ge1$ and $\epsilon\in(0,\frac1{16}]$, and consider either variant.
\begin{enumerate}[label=(\roman*)]
\item A deterministic algorithm with at most $\frac{32}{3\ln2}\,\epsilon^2N+3\log_2(N+2)+3$ bits of advice has ratio at least $\frac34+\epsilon$ on every instance.
\item Every randomized algorithm whose expected ratio is at least $\frac34+\epsilon$ on every instance reads more than $\frac{\epsilon^2}{\ln2}N-\log_2\frac2\epsilon$ bits of advice.
\end{enumerate}
Hence there is a constant $C$ such that the advice complexity of ratio $\frac34+\epsilon$ is $\Theta(\epsilon^2N)$ whenever $\epsilon^2N\ge C\log N$, for deterministic and for randomized algorithms.
\end{corollary}

\noindent The proof is in Appendix~\ref{app:hidden}. The lower bound uses only concentration and the upper bound only covering, so the advice complexity of the family is a covering number. For the high-bid variant with $|S|=t$ targets this can be made exact. We call $\mathcal R$ a \emph{$\lambda$-defect covering family} for $(N,t)$ if every $t$-set $S$ has some $R\in\mathcal R$ with $|S\cap R|\ge\max\{|R|,t\}-\lambda$, and we let $M(N,t,\lambda)$ be the least size of such a family.

\begin{theorem}
\label{thm:char}
For variant $\mathsf H$ with $t$ targets and Phase-1 demand $(N-t)B$, the least advice for deterministic ratio $\rho$ is exactly $\lceil\log_2M(N,t,\lfloor(1-\rho)N\rfloor)\rceil$.
\end{theorem}

\noindent The proof and estimates of $M$ are in Appendix~\ref{app:hidden}. For a margin below $t/8$ and $t\le\sqrt N$ we get $\log_2M=\Theta(t\log\frac Nt)$, so the advice must essentially name $S$. At margin $\Theta(t)$ it collapses to $\Theta(\epsilon^2N)$.

\section{Further Results}
\label{sec:further}

\paragraph{Ratios just above $1-1/e$.} On independent blocks of the nested family, ratio $1-\frac1e+\delta$ costs $\Omega(\delta^3n)$ bits (Theorem~\ref{thm:balance-barrier} in Appendix~\ref{app:further}). One bit per advertiser already lifts the ratio on the nested family to about $0.734$ (Appendix~\ref{app:priority}).

\paragraph{All unit-bid instances.} With one proportional weight per advertiser \cite{agrawal2018proportional} as advice, $2n\log_2\frac1\epsilon+n\log_2\log\frac n\epsilon+O(n)$ bits give ratio $1-\epsilon$ on \emph{every} divisible unit-bid instance (Theorem~\ref{thm:unit}). Hence $\Theta(n\log\frac1\epsilon)$ is the answer beyond decreasing instances whenever $\frac1n\le\epsilon\le\frac1{\log n}$.

\paragraph{Large bids.} Without the small-bid assumption no deterministic algorithm without advice is competitive. One bit gives ratio $1/6$ on every instance (Theorem~\ref{thm:one-bit}). For large $m/n$, every ratio above $1/2$ costs $\Omega(n\log\frac mn)$ bits even when every query has a single eligible advertiser (Theorem~\ref{thm:half-barrier}). In that case $O(\frac n\epsilon\log m)$ bits give ratio $1-2\epsilon$ (Theorem~\ref{thm:private}). For one advertiser these are known facts about online knapsack \cite{bockenhauer2012advice}, and our contribution is the extension to many advertisers. Exact optimality costs $m\log_2(\Delta+1)$ bits up to lower-order terms, for $m$ queries of degree at most $\Delta$ (Theorem~\ref{thm:exact-large}). Since an $\mathrm{AOC}$-complete problem needs $\Omega(\ell)$ bits for ratio $1/6$ on inputs of length $\ell$ \cite{boyar2015advice}, no class of OAW-HB instances is $\mathrm{AOC}$-complete.

\section{Conclusion}
\label{sec:conclusion}

Three questions remain open. (1) Do $n\log_2\frac1\epsilon+O(n)$ bits suffice on all small-bid instances, and is $-n\log_2e$ the right second-order term on every decreasing instance? For matching, where fractional solutions cannot be rounded online, the bounds are still $\Omega(n\log\frac1\epsilon)$ and $O(n/\epsilon^5)$ \cite{durr2016power}. (2) Is $\Theta(\epsilon^2n)$ the answer for the whole two-phase class? (3) For large bids and private queries, the cost of ratio $1-\epsilon$ is known only up to a factor of order $1/\epsilon$.

\bibliographystyle{splncs04}
\bibliography{ref}

@inproceedings{bockenhauer2009advice,
  author    = {B{\"o}ckenhauer, Hans-Joachim and Komm, Dennis and Kr{\'a}lovi{\v{c}}, Rastislav and Kr{\'a}lovi{\v{c}}, Richard and M{\"o}mke, Tobias},
  title     = {On the Advice Complexity of Online Problems},
  booktitle = {International Symposium on Algorithms and Computation (ISAAC)},
  series    = {Lecture Notes in Computer Science},
  volume    = {5878},
  pages     = {331--340},
  publisher = {Springer},
  year      = {2009}
}

@inproceedings{bockenhauer2012advice,
  author    = {B{\"o}ckenhauer, Hans-Joachim and Komm, Dennis and Kr{\'a}lovi{\v{c}}, Richard and Rossmanith, Peter},
  title     = {On the Advice Complexity of the Knapsack Problem},
  booktitle = {Latin American Symposium on Theoretical Informatics (LATIN)},
  series    = {Lecture Notes in Computer Science},
  volume    = {7256},
  pages     = {61--72},
  publisher = {Springer},
  year      = {2012}
}

@article{bockenhauer2014string,
  author  = {B{\"o}ckenhauer, Hans-Joachim and Hromkovi{\v{c}}, Juraj and Komm, Dennis and Krug, Sacha and Smula, Jasmin and Sprock, Andreas},
  title   = {The String Guessing Problem as a Method to Prove Lower Bounds on the Advice Complexity},
  journal = {Theoretical Computer Science},
  volume  = {554},
  pages   = {95--108},
  year    = {2014}
}

@article{boyar2017online,
  author  = {Boyar, Joan and Favrholdt, Lene M. and Kudahl, Christian and Larsen, Kim S. and Mikkelsen, Jesper W.},
  title   = {Online Algorithms with Advice: A Survey},
  journal = {ACM Computing Surveys},
  volume  = {50},
  number  = {2},
  pages   = {1--34},
  year    = {2017}
}

@inproceedings{dobrev2008much,
  author    = {Dobrev, Stefan and Kr{\'a}lovi{\v{c}}, Rastislav and Pardubsk{\'a}, Dana},
  title     = {How Much Information about the Future Is Needed?},
  booktitle = {International Conference on Current Trends in Theory and Practice of Computer Science (SOFSEM)},
  series    = {Lecture Notes in Computer Science},
  volume    = {4910},
  pages     = {247--258},
  publisher = {Springer},
  year      = {2008}
}

@inproceedings{durr2016power,
  author    = {D{\"u}rr, Christoph and Konrad, Christian and Renault, Marc},
  title     = {On the Power of Advice and Randomization for Online Bipartite Matching},
  booktitle = {24th Annual European Symposium on Algorithms (ESA 2016)},
  series    = {Leibniz International Proceedings in Informatics (LIPIcs)},
  volume    = {57},
  pages     = {37:1--37:16},
  year      = {2016},
  doi       = {10.4230/LIPIcs.ESA.2016.37}
}

@inproceedings{hromkovivc2010information,
  author    = {Hromkovi{\v{c}}, Juraj and Kr{\'a}lovi{\v{c}}, Rastislav and Kr{\'a}lovi{\v{c}}, Richard},
  title     = {Information Complexity of Online Problems},
  booktitle = {International Symposium on Mathematical Foundations of Computer Science (MFCS)},
  series    = {Lecture Notes in Computer Science},
  volume    = {6281},
  pages     = {24--36},
  publisher = {Springer},
  year      = {2010}
}

@article{mehta2007adwords,
  author  = {Mehta, Aranyak and Saberi, Amin and Vazirani, Umesh and Vazirani, Vijay},
  title   = {{AdWords} and Generalized Online Matching},
  journal = {Journal of the ACM},
  volume  = {54},
  number  = {5},
  pages   = {22},
  year    = {2007}
}

@article{miyazaki2014advice,
  author  = {Miyazaki, Shuichi},
  title   = {On the Advice Complexity of Online Bipartite Matching and Online Stable Marriage},
  journal = {Information Processing Letters},
  volume  = {114},
  number  = {12},
  pages   = {714--717},
  year    = {2014}
}

@inproceedings{mikkelsen2016randomization,
  author    = {Mikkelsen, Jesper W.},
  title     = {Randomization Can Be as Helpful as a Glimpse of the Future in Online Computation},
  booktitle = {43rd International Colloquium on Automata, Languages, and Programming (ICALP 2016)},
  series    = {LIPIcs},
  volume    = {55},
  pages     = {39:1--39:14},
  year      = {2016}
}

@misc{pena2016limitations,
  author = {Pena, Nicolas and Borodin, Allan},
  title  = {On the Limitations of Deterministic De-randomizations for Online Bipartite Matching and Max-{SAT}},
  year   = {2016},
  note   = {arXiv:1608.03182}
}

@inproceedings{buchbinder2023lossless,
  author    = {Buchbinder, Niv and Naor, Joseph and Wajc, David},
  title     = {Lossless Online Rounding for Online Bipartite Matching (Despite its Impossibility)},
  booktitle = {Proceedings of the 2023 ACM-SIAM Symposium on Discrete Algorithms (SODA)},
  year      = {2023},
  doi       = {10.1137/1.9781611977554.ch78}
}

@article{hoeffding1963,
  author  = {Hoeffding, Wassily},
  title   = {Probability Inequalities for Sums of Bounded Random Variables},
  journal = {Journal of the American Statistical Association},
  volume  = {58},
  number  = {301},
  pages   = {13--30},
  year    = {1963}
}

@inproceedings{boyar2015advice,
  author    = {Boyar, Joan and Favrholdt, Lene M. and Kudahl, Christian and Mikkelsen, Jesper W.},
  title     = {Advice Complexity for a Class of Online Problems},
  booktitle = {32nd International Symposium on Theoretical Aspects of Computer Science (STACS 2015)},
  series    = {LIPIcs},
  volume    = {30},
  pages     = {116--129},
  year      = {2015}
}

@article{emek2011online,
  author  = {Emek, Yuval and Fraigniaud, Pierre and Korman, Amos and Ros{\'e}n, Adi},
  title   = {Online Computation with Advice},
  journal = {Theoretical Computer Science},
  volume  = {412},
  number  = {24},
  pages   = {2642--2656},
  year    = {2011}
}

@inproceedings{agrawal2018proportional,
  author    = {Agrawal, Shipra and Zadimoghaddam, Morteza and Mirrokni, Vahab},
  title     = {Proportional Allocation: Simple, Distributed, and Diverse Matching with High Entropy},
  booktitle = {Proceedings of the 35th International Conference on Machine Learning (ICML)},
  series    = {Proceedings of Machine Learning Research},
  volume    = {80},
  pages     = {99--108},
  year      = {2018}
}

\appendix
\section{Priority Classes, Exact Losses and Indivisible Bids}
\label{app:priority}

\paragraph{The divisible model.} We work in the small-bid (fluid) limit. Advertiser $i$ has budget $B_i$, demand arrives as a sequence of parcels, a parcel of mass $\mu$ with eligible set $N$ may be split arbitrarily among the advertisers of $N$, and no advertiser is loaded beyond its budget. This is the limit of instances with small bids, and the model in which BALANCE is exactly $(1-1/e)$-competitive \cite{mehta2007adwords}.

\paragraph{A demand-based variant.} Theorem~\ref{thm:pc-all} compares the algorithm with an optimal solution. The following variant compares it with the total demand instead, and for this reason it extends to indivisible bids. For a decreasing instance we partition the rounds into consecutive \emph{windows} $W_1,\dots,W_L$. Class $a$ consists of the advertisers with $\tau_i\in W_a$ (advertisers that are never eligible play no role), $C_a$ is its total budget and $D_a$ is the demand of window $a$, and we put $C=\sum_aC_a$ and $D=\sum_aD_a$. The windows are \emph{equitable} if $C_1=\dots=C_L=C/L$, and \emph{balanced} if $D_a\le C_a$ for every $a$. In the indivisible version every query $q$ has a single bid $b_q$, the same for all advertisers eligible for it. An advertiser \emph{has room} for $q$ if its remaining budget is at least $b_q$, and $\mathrm{BAL}_L$ assigns $q$ to an eligible advertiser with room that minimizes (class, fill fraction) lexicographically, or rejects $q$ if there is none. For $\gamma\in[0,1)$ the instance is \emph{$\gamma$-small} if $b_q\le\gamma B_i$ for every query $q$ and every advertiser $i$ eligible for it. The divisible model is the case $\gamma=0$.

\begin{theorem}
\label{thm:decreasing}
Let a decreasing instance be divisible ($\gamma=0$) or $\gamma$-small with indivisible queries, and let the windows satisfy $D_a\le(1-\gamma)C_{a+1}$ for every $a<L$. Then $\mathrm{BAL}_L$, with $n\lceil\log_2L\rceil$ bits of advice, rejects only demand of window $L$, so it serves at least $D-D_L$. In particular, if the instance is divisible and the windows are equitable and balanced, then $\mathrm{BAL}_L\ge D-\frac CL$, and its ratio is at least $1-\frac1L$ when $D=C$. If the instance is $\gamma$-small and the windows are equitable with $D_a\le(1-\gamma)\frac CL$ for every $a$, then $\mathrm{BAL}_L\ge D-(1-\gamma)\frac CL$.
\end{theorem}
\begin{proof}
Every advertiser $i$ is eligible exactly in the rounds $t\le\tau_i$, because $i\in E_{\tau_i}\subseteq E_t$ for $t\le\tau_i$. Hence during window $a$ the advertisers of the classes below $a$ have exited and those of the classes above $a$ are all live. We show by induction on $a$ that during window $a$ all demand is assigned to the classes $a$ and $a+1$, and that none is rejected if $a<L$.

Fix $a<L$ and assume the claim for the earlier windows. Then no demand has reached class $a+1$ before window $a$. During window $a$ the algorithm prefers class $a$. It moves on to class $a+1$ only when no live member of class $a$ has room, and beyond class $a+1$, or to a rejection, only when in addition no member of class $a+1$ has room. Suppose that at some moment of window $a$ no member of class $a+1$ has room for the current demand. In the divisible model every member of class $a+1$ is then full, so the load of the class is $C_{a+1}\ge D_a$. In the indivisible model the current query $q$ finds every member $i$ of class $a+1$ with remaining budget below $b_q\le\gamma B_i$, so the load of the class exceeds $(1-\gamma)C_{a+1}\ge D_a$. But all of this load comes from window $a$, whose total demand is $D_a$ and includes the current demand, which is not yet placed. So the load is less than $D_a$, a contradiction. Hence during window $a$ every demand that class $a$ does not take goes to class $a+1$, and nothing is rejected.

Consequently, all demand of the windows $1,\dots,L-1$ is served, the rejected demand is at most $D_L$, and $\mathrm{BAL}_L\ge D-D_L$. For the two special cases note that $D_a\le C_a=C_{a+1}$ and $D_L\le\frac CL$ in the first, and $D_a\le(1-\gamma)C_{a+1}$ and $D_L\le(1-\gamma)\frac CL$ in the second. \qed
\end{proof}

\paragraph{The family.} Let $\mathcal N(g)$ consist of the following instances. There are $g$ advertisers of budget $1$ and a hidden bijection $\pi$ from positions $[g]$ to advertisers. Advertiser $\pi(p)$ is eligible in the phases $1,\dots,p$ and \emph{exits} after phase $p$. Phase $t$ is a parcel of mass $1$ with eligible set $E_t=\{\pi(t),\dots,\pi(g)\}$, so $E_1\supsetneq E_2\supsetneq\cdots\supsetneq E_g$ is a laminar chain. Assigning phase $t$ to $\pi(t)$ serves every parcel and exhausts every budget, so $\mathrm{OPT}=g$ and the total budget equals the total demand.

\paragraph{The algorithm.} The advice is a map $\ell$ from the advertisers to $[L]$, which costs $g\lceil\log_2L\rceil$ bits. Algorithm $\mathrm{BAL}_L$ routes the arriving demand to an eligible advertiser that is not full and minimizes $(\ell_i,f_i)$ lexicographically, where $f_i$ is the fraction of $B_i$ already spent. Equivalently, for each parcel let $a$ be the smallest class containing an eligible non-full advertiser, raise the common fill fraction of the eligible non-full advertisers of class $a$ until the parcel is exhausted or they are full, and in the latter case continue with the next class. For $L=1$ this is BALANCE.

\begin{theorem}
\label{thm:priority}
Let $L$ divide $g$, put $h=g/L$, and let the oracle set $\ell_{\pi(p)}=\lceil pL/g\rceil$. Then on every instance of $\mathcal N(g)$,
\[
\mathrm{BAL}_L\ \ge\ \Big(1-\frac1L\Big)\,\mathrm{OPT},
\]
with $g\lceil\log_2L\rceil$ bits of advice.
\end{theorem}
\begin{proof}
We call \emph{window} $a$ the phases $(a-1)h+1,\dots,ah$ and \emph{class} $a$ the $h$ advertisers whose positions lie in window $a$. Class $a$ has total budget $h$, and all its advertisers exit during window $a$. Let $f_i$ be the fill fraction of $i$ when it exits and $W=\sum_i(1-f_i)$ the total budget left unused. Since the budgets are $1$ and the total demand equals the total budget $g$,
\begin{equation}\label{eq:pc-rev}
\mathrm{BAL}_L=\sum_if_i=g-W .
\end{equation}
We prove $W\le h$, which together with \eqref{eq:pc-rev} gives the claim.

We show by induction on $a$ that at the beginning of window $a$ (i) the advertisers of class $a$ are all live and share a common fill fraction $\rho_a<1$, (ii) every advertiser of a class above $a$ has fill $0$, and (iii) every advertiser of a class below $a$ has exited. For $a=1$ this holds with $\rho_1=0$.

Assume (i)--(iii) for $a$. By (iii) the classes below $a$ are gone, so class $a$ is the lowest class with live advertisers for the whole of window $a$, and $\mathrm{BAL}_L$ routes demand to class $a+1$ only when the live members of class $a$ are full. Within window $a$ the live members of class $a$ keep a common fill, because by (i) they start equal, the rule raises the common fill of all of them, and the departure of one does not change the fill of the others. We write $\lambda_0=\rho_a$ and $\lambda_j$ for their common fill after the $j$-th phase of the window. Exactly $h-j+1$ of them are live during phase $j$, so
\begin{equation}\label{eq:pc-wf}
\lambda_j=\min\Big\{1,\ \lambda_{j-1}+\frac1{h-j+1}\Big\},
\end{equation}
and the member exiting after phase $j$ does so with fill $\lambda_j$. Let $S_a$ be the total mass routed out of class $a$ during window $a$. By (ii) the recipients are the members of class $a+1$, which are all live and all at fill $0$ at the start of the window, so the rule keeps their fills equal, and we put $\rho_{a+1}=S_a/h$. In phase $1$ of the window the $h$ members of class $a$ are live with fill $\rho_a<1$, so class $a$ absorbs at least $\min\{1,h(1-\rho_a)\}>0$ of that phase, whence
\[
S_a\ \le\ h-\min\{1,h(1-\rho_a)\}\ <\ h ,
\]
so $\rho_{a+1}<1$. In particular the members of class $a+1$ never become full during window $a$, so no mass reaches a class beyond $a+1$ and none is rejected, and (i)--(iii) hold for $a+1$.

It remains to compute $W$. Class $a$ receives $\rho_ah$ during window $a-1$ and $h-S_a$ during window $a$, since the mass $h$ of window $a$ is routed entirely to the classes $a$ and $a+1$, and it receives nothing at any other time. Hence $\sum_{i\in\text{class }a}f_i=\rho_ah+h-S_a$ and
\[
W_a:=\sum_{i\in\text{class }a}(1-f_i)=h-\rho_ah-h+S_a=h\,(\rho_{a+1}-\rho_a),
\]
where for $a=L$ we read $\rho_{L+1}=S_L/h$, with $S_L<h$ the mass of window $L$ that class $L$ does not absorb (it is lost, since there is no class $L+1$). Summing over $a$ telescopes to
\[
W=\sum_{a=1}^LW_a=h\,(\rho_{L+1}-\rho_1)=h\,\rho_{L+1}\ <\ h . \qquad\qed
\]
\end{proof}

The proof also gives the exact loss $h\rho_{L+1}$, and \eqref{eq:pc-wf} determines $\rho_{L+1}$.

\begin{proposition}
\label{prop:pc-exact}
Fix $L$ and let $h\to\infty$. Then $\rho_{a}\to r_a$, where $r_1=0$ and $r_{a+1}=e^{r_a-1}$, and the competitive ratio of $\mathrm{BAL}_L$ on $\mathcal N(g)$ tends to $1-r_{L+1}/L$. Moreover $1-r_{L+1}=\Theta(1/L)$, so the bound of Theorem~\ref{thm:priority} is tight up to its second-order term.
\end{proposition}
\begin{proof}
By \eqref{eq:pc-wf}, as long as $\lambda_j<1$ we have $\lambda_j=\rho_a+(H_h-H_{h-j})$, with $H$ the harmonic numbers. Let $j^*$ be the least $j$ with $\rho_a+H_h-H_{h-j}\ge1$. The phases after $j^*$ contribute their whole mass to class $a+1$ and phase $j^*$ contributes a part, so $S_a=h-j^*+O(1)$ and $\rho_{a+1}=(h-j^*)/h+O(1/h)$. Since $H_h-H_{h-j}=\ln\frac h{h-j}+O\big(\frac1{h-j}\big)$, and $h-j^*\ge h/e-O(1)$ because $\rho_a\ge0$, we get $(h-j^*)/h=e^{\rho_a-1}+O(1/h)$. The map $\rho\mapsto e^{\rho-1}$ is continuous and maps $[0,1)$ into itself, and $L$ is fixed, so composing $L$ windows gives $\rho_{L+1}\to r_{L+1}$, and the ratio $1-\rho_{L+1}/L$ tends to $1-r_{L+1}/L$. Finally, with $x_a=1-r_a$ the recursion reads $x_{a+1}=1-e^{-x_a}=x_a-\frac{x_a^2}2+O(x_a^3)$, whence $x_a=(2+o(1))/a$. \qed
\end{proof}

\section{Estimates for Theorem~\ref{thm:nested-ub}}
\label{app:nested}

We bound $\ln\frac1P$ for the probability $P$ in the proof of Theorem~\ref{thm:nested-ub}. In the first product, $t/(n-t+1)\le\frac13$ and $-\ln(1-x)\le\frac32x$ give a contribution of at most $K(K+1)/n$ to $\ln\frac1P$. For the second, $N!\le eN^{N+1/2}e^{-N}$ and $N!\ge(N/e)^N$ give
\begin{align*}
\ln\frac{(n-K)!}{(K+1)!\,(K+1)^{n-2K-1}}&\le(n-K)\ln\frac{n-K}{K+1}-n+2K+2+\tfrac12\ln n\\
&\le n\ln\frac n{e(K+1)}-K\ln\frac n{K+1}+2K+2+\tfrac12\ln n.
\end{align*}
Together with the bound $\log_2(1/P)+2\log_2n+1$ on the advice this gives the theorem.

\section{Proofs for Section~\ref{sec:hidden}}
\label{app:hidden}

\begin{lemma}
\label{lem:anticonc}
Let $k$ be even, $t$ an integer with $0\le t\le k/4$, $S$ a fixed $k$-subset of $[2k]$, and $T$ a uniformly random $k$-subset. Then
\[
\Pr\big[|S\cap T|=\tfrac k2-t\big]\ \ge\ \frac{1}{(k+1)^2}\exp\Big(-\frac{16t^2}{3k}\Big).
\]
\end{lemma}
\begin{proof}
With $j=\frac k2-t$, choosing $T$ amounts to choosing $j$ elements of $S$ and $k-j$ elements of its complement, so $\Pr[|S\cap T|=j]=\binom kj\binom k{k-j}/\binom{2k}k=\binom kj^2/\binom{2k}k$. Let $H$ be the binary entropy and $p=j/k$. The term $\binom kjp^j(1-p)^{k-j}$ is the largest of the $k+1$ terms of the binomial distribution with parameters $k$ and $p$, so it is at least $\frac1{k+1}$. Since $p^j(1-p)^{k-j}=2^{-kH(p)}$, this gives $\binom kj\ge2^{kH(j/k)}/(k+1)$. Together with $\binom{2k}k\le4^k$ we obtain $\Pr\ge2^{-2k(1-H(1/2-x))}/(k+1)^2$ with $x=t/k\le1/4$. From the expansion
\[
1-H\big(\tfrac12-x\big)=\frac1{\ln2}\sum_{r\ge1}\frac{(2x)^{2r}}{2r(2r-1)}\ \le\ \frac1{2\ln2}\cdot\frac{4x^2}{1-4x^2}\ \le\ \frac{8x^2}{3\ln2}\qquad(x\le\tfrac14)
\]
we get $2^{-2k(1-H(1/2-x))}\ge\exp(-\frac{16}3kx^2)=\exp(-\frac{16t^2}{3k})$. \qed
\end{proof}

\begin{proof}[of Theorem~\ref{thm:hidden-upper}]
\emph{The algorithm.} The advice names a \emph{reservation set} $R$ with $|R|=k$ from a fixed list. In Phase~1 the algorithm gives $B$ queries to each advertiser of $R$, which uses all $kB$ queries and earns $kB$. In Phase~2 it accepts every feasible query. The advertisers outside $R$ are untouched, so in either variant every target outside $R$ earns $B$. Hence, if $|S\cap R|\le\frac k2-t$,
\[
\mathrm{ALG}\ \ge\ kB+B\,|S\setminus R|\ \ge\ kB+\Big(\frac k2+t\Big)B,
\]
and since $\mathrm{OPT}=2kB$ the ratio is at least $\frac34+\frac t{2k}=\frac34+\frac tN$.

\emph{The list.} We say that $R$ serves $S$ if $|S\cap R|\le\frac k2-t$. For a uniformly random $k$-set $R$, Lemma~\ref{lem:anticonc} gives
\[
\Pr[R\text{ serves }S]\ \ge\ \Pr\Big[|S\cap R|=\frac k2-t\Big]=\binom k{k/2-t}^2\Big/\binom{2k}k\ \ge\ q:=\frac{e^{-16t^2/(3k)}}{(k+1)^2}
\]
for every fixed $S$. Lemma~\ref{lem:cover}, applied to the $|\mathcal S|=\binom{2k}k\le4^k$ target sets, yields a list of at most $\ln(4^k)/q+1=2k\ln2\,(k+1)^2e^{16t^2/(3k)}+1\le M$ sets that serves every $S$, where we used $(k+1)^2\ln2\ge1$. The oracle writes the index of a serving set. \qed
\end{proof}

\begin{proof}[of Corollary~\ref{cor:hidden}]
(i) We take $t=\lceil\epsilon N\rceil$ in Theorem~\ref{thm:hidden-upper} and check three estimates.
\begin{itemize}
\item \emph{Range and ratio.} $t\le2\epsilon k+1\le\frac k8+1\le\frac k4$ because $k\ge8$, and the ratio is at least $\frac34+\frac tN\ge\frac34+\epsilon$.
\item \emph{Exponent.} $\frac{16t^2}{3k}\le\frac{16}3\big(4\epsilon^2k+4\epsilon+\frac1k\big)\le\frac{32}3\epsilon^2N+2$, using $4\epsilon\le\frac14$ and $\frac1k\le\frac18$.
\item \emph{Polynomial factor.} $(2k+1)(k+1)^2\le\frac{(N+2)^3}4$.
\end{itemize}
With $x=(2k+1)(k+1)^2\ln2\cdot e^{16t^2/(3k)}\ge1$, the advice is $\lceil\log_2\lceil x\rceil\rceil\le\log_2x+2$, and
\begin{align*}
\log_2x\ &\le\ 3\log_2(N+2)-2+\log_2\ln2+\Big(\frac{32}3\epsilon^2N+2\Big)\log_2e\\
&\le\ \frac{32}{3\ln2}\epsilon^2N+3\log_2(N+2)+1 .
\end{align*}
(ii) If the advice were at most $\frac{\epsilon^2}{\ln2}N-\log_2\frac2\epsilon$ bits, then by Theorem~\ref{thm:hidden} some instance would have expected ratio at most $\frac34+\frac58\epsilon<\frac34+\epsilon$.

For the last claim, $\epsilon^2N\ge C\log N$ makes both $3\log_2(N+2)+3$ and $\log_2\frac2\epsilon\le1+\frac12\log_2N$ at most a small constant fraction of $\epsilon^2N$. \qed
\end{proof}

Recall that $\mathcal R$ is a \emph{$\lambda$-defect covering family} for $(N,t)$ if for every $S\subseteq[N]$ with $|S|=t$ there is $R\in\mathcal R$ with
\begin{equation}\label{eq:defect}
|S\cap R|\ \ge\ \max\{|R|,t\}-\lambda .
\end{equation}
\begin{lemma}
\label{lem:usable}
If $R$ satisfies \eqref{eq:defect} for some $S$ with $|S|=t$, then $|R|\le t+\lambda$.
\end{lemma}
\begin{proof}
$|R|-\lambda\le\max\{|R|,t\}-\lambda\le|S\cap R|\le|S|=t$. \qed
\end{proof}

\begin{proof}[of Theorem~\ref{thm:char}]
Fix a deterministic algorithm and an advice string $\sigma$, and let $R_\sigma$ be the set of advertisers on which it spends nothing during the first phase. Since the first phase is the same in all instances of the family, $R_\sigma$ does not depend on $S$ (execution invariance). The first phase earns the total spend, which is at most the capacity $B(N-|R_\sigma|)$ of the advertisers it touches and at most the demand $(N-t)B$. In the second phase a query of bid $B$ is accepted only by an advertiser whose budget is untouched, that is, by a member of $S\cap R_\sigma$. Hence
\begin{equation}\label{eq:char-alg}
\mathrm{ALG}\ \le\ B\min\{N-t,\ N-|R_\sigma|\}+B\,|S\cap R_\sigma| ,
\end{equation}
and, since $\mathrm{OPT}=NB$, the algorithm is $\rho$-competitive on the instance $X(S)$ only if
\[
\min\{N-t,N-|R_\sigma|\}+|S\cap R_\sigma|\ \ge\ \rho N .
\]
The left-hand side is an integer and $\lceil\rho N\rceil=N-\lfloor(1-\rho)N\rfloor=N-\lambda$, so the displayed inequality is equivalent to \eqref{eq:defect} for $R_\sigma$. Therefore a $\rho$-competitive algorithm that reads at most $b$ bits yields a $\lambda$-defect covering family $\{R_\sigma:\sigma\in\{0,1\}^b\}$, whence $2^b\ge M(N,t,\lambda)$.

Conversely, let $\mathcal R$ be a $\lambda$-defect covering family of size $M$. The oracle writes, in $\lceil\log_2M\rceil$ bits, the index of some $R\in\mathcal R$ that satisfies \eqref{eq:defect} for the hidden $S$. The algorithm accepts a first-phase query whenever some advertiser outside $R$ still has budget, and assigns it there, and it accepts every second-phase query it can. The first phase then earns exactly $B\min\{N-t,N-|R|\}$ and the second at least $B|S\cap R|$, so $\mathrm{ALG}\ge B\min\{N-t,N-|R|\}+B|S\cap R|$, and by \eqref{eq:defect} this is at least $B(N-\lambda)\ge\rho NB$. \qed
\end{proof}

\begin{theorem}
\label{thm:Mvalues}
Let $N\ge t\ge1$ and $\lambda\ge0$ be integers.
\begin{enumerate}[label=(\roman*)]
\item $M(N,t,0)=\binom Nt$, and $M(N,t,\lambda)\le\binom Nt$ for every $\lambda$.
\item If $\lambda\le t/3$ then
$\displaystyle M(N,t,\lambda)\ \ge\ \binom Nt\Big/\Big[\binom{t+\lambda}{t-\lambda}\binom N\lambda\Big].$
\item If $1\le\lambda\le t/8$, $t\le\sqrt N$ and $N\ge64$, then
\[
\tfrac14\,t\log_2\tfrac Nt\ \le\ \log_2M(N,t,\lambda)\ \le\ t\log_2\tfrac{eN}t ,
\]
so $\log_2M=\Theta\big(t\log\frac Nt\big)=\Theta\big(\log\binom Nt\big)$.
\end{enumerate}
\end{theorem}
\begin{proof}
(i) The family of all $t$-subsets satisfies \eqref{eq:defect} with $R=S$ for every $\lambda\ge0$. For $\lambda=0$, condition \eqref{eq:defect} forces $S\subseteq R$ and $|R|\le t$, so a usable $R$ equals the unique $S$ it serves, and no family smaller than $\binom Nt$ works.

(ii) Fix a usable $R$, and put $r=|R|\le t+\lambda$ (Lemma~\ref{lem:usable}) and $q=\max\{r,t\}-\lambda$. The sets $S$ that $R$ serves have $|S\cap R|\ge q$, so each is obtained by choosing $q$ of its elements in $R$ and the remaining $t-q$ anywhere, which gives at most $\binom rq\binom N{t-q}$ of them. Now $t-q=t-\max\{r,t\}+\lambda\le\lambda\le N/2$, so $\binom N{t-q}\le\binom N\lambda$. Also $q\ge t-\lambda$ and $r\le t+\lambda$, so $\binom rq\le\binom{t+\lambda}q$, and since $t\ge3\lambda$ we have $t-\lambda\ge(t+\lambda)/2$. Hence $q\mapsto\binom{t+\lambda}q$ is non-increasing on $[t-\lambda,t+\lambda]$ and $\binom rq\le\binom{t+\lambda}{t-\lambda}$. A family must serve all $\binom Nt$ sets $S$, which gives the bound.

(iii) The upper bound is (i) together with $\binom Nt\le(eN/t)^t$. For the lower bound we take logarithms in (ii) and use $\binom Nt\ge(N/t)^t$, $\binom N\lambda\le(eN/\lambda)^\lambda$ and $\binom{t+\lambda}{t-\lambda}\le\big(\frac{e(t+\lambda)}{2\lambda}\big)^{2\lambda}$, which give
\[
\log_2M\ \ge\ t\log_2\tfrac Nt-\lambda\log_2\tfrac{eN}\lambda-2\lambda\log_2\tfrac{e(t+\lambda)}{2\lambda} .
\]
Since $t\le\sqrt N$ we have $\log_2\frac Nt\ge\frac12\log_2N$ and $\log_2t\le\frac12\log_2N\le\log_2\frac Nt$. Using $\lambda\le t/8$, the second term is at most $\frac t8(\log_2N+\log_2e)\le\frac14t\log_2\frac Nt+\frac t5$, and the third is at most $\frac t4\log_2\frac{9et}{16\lambda}\le\frac t4(\log_2t+1)\le\frac14t\log_2\frac Nt+\frac t4$. As $N/t\ge\sqrt N\ge8$ we have $\log_2\frac Nt\ge3$, so $\frac t5+\frac t4\le\frac14t\log_2\frac Nt$, and the three subtracted terms leave at least $\frac14t\log_2\frac Nt$. \qed
\end{proof}

\section{Proofs for Section~\ref{sec:further}}
\label{app:further}

\subsection{The Barrier at $1-1/e$}

\paragraph{The family.} Let $B\ge1$, $g\ge3$ and $s\ge1$ be integers and $n=gs$. There are $s$ blocks, and block $r$ has $g$ advertisers $u_{r,1},\dots,u_{r,g}$ of budget $B$ and a permutation $\pi_r\in S_g$. Block $r$ consists of $g$ \emph{phases}, and phase $t$ is a run of $B$ queries, each bidding $1$ on every advertiser of $N_{r,t}=\{u_{r,\pi_r(h)}:t\le h\le g\}$. Blocks are presented one after the other. Every bid is the fraction $1/B$ of a budget, and assigning phase $t$ of block $r$ to $u_{r,\pi_r(t)}$ serves every query, so $\mathrm{OPT}=gsB$. The neighborhoods $N_{r,1}\supset N_{r,2}\supset\dots\supset N_{r,g}$ of the phases of one block form a laminar chain.

\begin{lemma}
\label{lem:block-loss}
Fix a deterministic algorithm (in particular, an algorithm with a fixed advice string) and fix everything presented before block $r$. If $\pi_r$ is uniform on $S_g$, the expected number of rejected queries of block $r$ is at least $B(g/e-2)$.
\end{lemma}

\begin{theorem}
\label{thm:balance-barrier}
Let $B\ge1$, $g\ge3$, $s\ge1$ and $n=gs$. Every deterministic algorithm that reads at most $b$ bits of advice has, on some instance of the family above (all bids equal to $1/B$ of the budget), competitive ratio at most
\[
1-\frac1e+\frac2g+\sqrt{\frac{b\ln2}{2s}} .
\]
Consequently, (i) with $b=0$ and $g,s\to\infty$, no advice-free deterministic algorithm has ratio above $1-1/e+o(1)$ on small-bid instances, and (ii) a ratio of at least $1-\frac1e+\delta$ with $\delta>\frac2g$ requires $b\ge\frac2{\ln2}\big(\delta-\frac2g\big)^2s$ bits. Choosing $g=\lceil4/\delta\rceil$ and $s=\lfloor n/g\rfloor$, with the remaining $n-gs$ advertisers receiving no queries, gives $b=\Omega(\delta^3n)$.
\end{theorem}
\begin{proof}[of Lemma~\ref{lem:block-loss}]
We drop the index $r$. For phase $j$ let $x_{j,u}$ be the number of its queries assigned to $u$ and $\rho_j=B-\sum_ux_{j,u}$ the number rejected. Fix $t\in\{2,\dots,g\}$ and let $T_t=\{u_{\pi(t)},\dots,u_{\pi(g)}\}$, of total capacity $(g-t+1)B$. Phases $j\ge t$ have neighborhoods inside $T_t$, and the mass $w_t=\sum_{j<t}\sum_{u\in T_t}x_{j,u}$ placed on $T_t$ by earlier phases is already spent. The $(g-t+1)B$ queries of phases $j\ge t$ can therefore be served only up to $(g-t+1)B-w_t$, so at least $w_t$ of them are rejected, and the number of rejected queries of the block is at least $\sum_{j<t}\rho_j+w_t$.

At phase $j$ the algorithm has seen only $\pi(1),\dots,\pi(j-1)$ (through the removed advertisers) and the earlier blocks, so $x_{j,\cdot}$ is determined by this history, while the advertisers of $N_{j}$ occupy the positions $j,\dots,g$ in a uniformly random order. Hence $\mathbb E[\sum_{u\in T_t}x_{j,u}\mid\text{history}]=(B-\rho_j)\frac{g-t+1}{g-j+1}$, and since $\rho_j+(B-\rho_j)q\ge Bq$ for $q\in[0,1]$,
\[
\mathbb E\Big[\sum_{j<t}\rho_j+w_t\Big]\ \ge\ B(g-t+1)\sum_{j<t}\frac1{g-j+1}=Bk(H_g-H_k),\qquad k=g-t+1 .
\]
We take $k=\lfloor g/e\rfloor$, so that $1\le k<g$ for $g\ge3$. Using $H_g-H_k\ge\ln\frac{g+1}{k+1}\ge\ln\frac{g+1}{g/e+1}\ge1-\frac{e-1}{g+1}$ and $k\ge g/e-1$, we get $k(H_g-H_k)\ge(g/e-1)\big(1-\frac{e-1}{g+1}\big)\ge g/e-2$. \qed
\end{proof}

\begin{proof}[of Theorem~\ref{thm:balance-barrier}]
For a string $\sigma\in\{0,1\}^b$ let $\mathcal A_\sigma$ be the advice-free algorithm that reads the tape $\sigma0^\infty$. An algorithm that reads at most $b$ bits behaves on every instance exactly as $\mathcal A_\sigma$ for the first $b$ bits $\sigma$ of its tape. Let $\pi=(\pi_1,\dots,\pi_s)$ be uniform and let $Y_r^\sigma$ be the number of rejected queries of block $r$ under $\mathcal A_\sigma$, so $Y_r^\sigma\in[0,gB]$ and $\mathcal L_\sigma(\pi)=\sum_rY_r^\sigma$ is the total loss. By Lemma~\ref{lem:block-loss}, $\mathbb E[Y_r^\sigma\mid\pi_1,\dots,\pi_{r-1}]\ge B(g/e-2)$, because $\pi_r$ is independent of the earlier blocks. Hence $\sum_r\big(Y_r^\sigma-\mathbb E[Y_r^\sigma\mid\pi_{<r}]\big)$ is a martingale whose increments lie in intervals of length $gB$, and the Azuma--Hoeffding inequality gives, for every $\lambda>0$,
\[
\Pr\big[\mathcal L_\sigma(\pi)\le sB(g/e-2)-\lambda\big]\le\exp\!\Big(-\frac{2\lambda^2}{s\,g^2B^2}\Big).
\]
Suppose the algorithm has loss at most $\ell$ on every instance. Then for every $\pi$ some $\sigma$ has $\mathcal L_\sigma(\pi)\le\ell$. If $\ell<sB(g/e-2)-\lambda$, a union bound over the $2^b$ strings gives $1\le2^b\exp(-2\lambda^2/(sg^2B^2))$, which fails as soon as $\lambda>gB\sqrt{sb\ln2/2}$. Hence $\ell\ge sB(g/e-2)-gB\sqrt{sb\ln2/2}$, and dividing by $\mathrm{OPT}=gsB$ gives the stated bound on the ratio. For (ii) we solve $\sqrt{b\ln2/(2s)}\ge\delta-2/g$, which is possible because $\delta-2/g>0$. With $g=\lceil4/\delta\rceil$ we have $\delta-2/g\ge\delta/2$ and $s\ge n\delta/5-1$. \qed
\end{proof}

\subsection{Large Bids}

\paragraph{Conventions.} An edge $(i,j)$ is \emph{usable} if $b_{ij}\le B_i$, and edges with $b_{ij}>B_i$ can never be served and are deleted. A usable edge is \emph{big} if $B_i/2<b_{ij}\le B_i$ and \emph{small} if $b_{ij}\le B_i/2$. Let $I_S$ ($I_B$) be the instance restricted to small (big) edges, with optimum $\mathrm{OPT}_S$ ($\mathrm{OPT}_B$). The restriction of an optimal allocation of $I$ to small (big) edges is feasible for $I_S$ ($I_B$), so
\begin{equation}\label{eq:split}
\mathrm{OPT}(I)\le\mathrm{OPT}_S+\mathrm{OPT}_B.
\end{equation}
We write $\rho_i$ for the remaining budget of advertiser $i$.

\begin{proposition}[Known for one advertiser \cite{bockenhauer2012advice}]
\label{prop:zero-advice}
Let $\mathcal A$ be a deterministic online algorithm that uses no advice, and let $\rho\in(0,1]$. There is an instance with one advertiser and at most two queries on which $\mathrm{ALG}<\rho\cdot\mathrm{OPT}$.
\end{proposition}
\begin{proof}
Let $B=1$ and $\delta=\rho/2$, and present a probe with bid $\delta$. If $\mathcal A$ accepts it, we present a second query (the commitment) with bid $1$. It is infeasible because $\delta+1>1$, so $\mathrm{ALG}=\rho/2<\rho=\rho\cdot\mathrm{OPT}$, as $\mathrm{OPT}=1$. If $\mathcal A$ rejects the probe, we stop, and then $\mathrm{ALG}=0<\rho\delta=\rho\cdot\mathrm{OPT}$. \qed
\end{proof}

\begin{lemma}
\label{lem:greedy-halves}
Let $\mathcal G_S$ (resp.\ $\mathcal G_B$) be the advice-free algorithm that, on each query $j$, considers only advertisers $i$ for which $(i,j)$ is a small (resp.\ big) usable edge with $b_{ij}\le\rho_i$, assigns $j$ to one of them with the largest bid (ties broken deterministically), and rejects $j$ if there is none. Then $\mathrm{ALG}(\mathcal G_S)\ge\mathrm{OPT}_S/3$ and $\mathrm{ALG}(\mathcal G_B)\ge\mathrm{OPT}_B/3$.
\end{lemma}
\begin{proof}
We run $\mathcal G_S$ and let $S_i$ be the final revenue from advertiser $i$ and $r_j$ the revenue collected on query $j$, so $\mathrm{ALG}=\sum_iS_i=\sum_jr_j$. We call $i$ \emph{full} if $S_i>B_i/2$. If $i$ is not full, then at every moment $\rho_i\ge B_i-S_i\ge B_i/2\ge b_{ij}$ for each small edge $(i,j)$, so $i$ is available for every query it has a small edge to. Let $M$ be an optimal allocation of $I_S$. For $(i,j)\in M$ with $i$ not full, $\mathcal G_S$ chose for $j$ an advertiser with bid at least $b_{ij}$, so $b_{ij}\le r_j$. Since $M$ uses each query at most once, these edges contribute at most $\sum_jr_j=\mathrm{ALG}$. For full $i$ the revenue of $M$ at $i$ is at most $B_i<2S_i$, so full advertisers contribute at most $2\,\mathrm{ALG}$. Hence $\mathrm{OPT}_S\le3\,\mathrm{ALG}$.

Two big bids of one advertiser exceed its budget, so in $\mathcal G_B$ and in every allocation of $I_B$ each advertiser serves at most one query. Let $M$ be an optimal allocation of $I_B$ and $(i,j)\in M$. If $i$ is unmatched at the end of $\mathcal G_B$, it was available when $j$ arrived, so $b_{ij}\le r_j$. If $i$ is matched with a bid $b'_i$, then $b'_i>B_i/2\ge b_{ij}/2$. The edges of the first kind involve distinct queries and contribute at most $\mathrm{ALG}$, and those of the second kind involve distinct advertisers and contribute less than $2\sum_ib'_i\le2\,\mathrm{ALG}$. Hence $\mathrm{OPT}_B\le3\,\mathrm{ALG}$. \qed
\end{proof}

\begin{theorem}[ Part (b) known for online knapsack \cite{bockenhauer2012advice}]
\label{thm:one-bit}
(a) There is a deterministic online algorithm that uses one bit of advice and is $1/6$-competitive on every OAW-HB instance, independently of $n$, $m$ and the bid sizes. (b) For $n=1$ one bit gives ratio $1/2$. (c) If every query has degree one, $n$ bits (one per advertiser) give ratio $1/2$.
\end{theorem}
\begin{proof}
(a) The oracle computes $\mathrm{OPT}_S$ and $\mathrm{OPT}_B$ and writes bit $0$ if $\mathrm{OPT}_S\ge\mathrm{OPT}_B$ and bit $1$ otherwise. The algorithm runs $\mathcal G_S$ on bit $0$ and $\mathcal G_B$ on bit $1$. By Lemma~\ref{lem:greedy-halves} and~\eqref{eq:split}, $\mathrm{ALG}\ge\max\{\mathrm{OPT}_S,\mathrm{OPT}_B\}/3\ge(\mathrm{OPT}_S+\mathrm{OPT}_B)/6\ge\mathrm{OPT}/6$.

(b) Let the bit be $1$ if and only if the instance contains a big usable item. On bit $1$ the algorithm accepts the first big item and rejects everything else, so $\mathrm{ALG}>B/2\ge\mathrm{OPT}/2$. On bit $0$ all usable items have size at most $B/2$, and the algorithm accepts each item that fits. If the final spend $S$ satisfies $S>B/2$, then $\mathrm{ALG}>B/2\ge\mathrm{OPT}/2$. Otherwise $S\le B/2$, so at every moment the remaining budget is at least $B/2$ and every item fit when it arrived. Hence all usable items were accepted and $\mathrm{ALG}\ge\mathrm{OPT}$.

(c) With degree-one queries the instance splits into $n$ independent single-advertiser instances, and each advertiser gets its own bit as in (b). Then $\mathrm{ALG}_i\ge\mathrm{OPT}_i/2$ for every $i$. \qed
\end{proof}

\begin{theorem}
\label{thm:half-barrier}
Let $\rho\in(1/2,1)$, $\theta=\frac{2\rho-1}{3-2\rho}\in(0,1)$, and let $n\ge1$ and $k\ge2$ be integers. There is a family $\mathcal F_{n,k}$ of OAW-HB instances with $n$ advertisers of equal budgets, $m=n(k+1)$ queries, and all queries of degree one, such that every deterministic algorithm that is $\rho$-competitive on $\mathcal F_{n,k}$ uses in the worst case at least
\[
c\log_2k-\log_2\binom nc\ \ge\ \theta n\log_2k-n\qquad\text{bits},\qquad c=\lceil\theta n\rceil.
\]
For $n=1$ the bound is $\lceil\log_2k\rceil=\lceil\log_2(m-1)\rceil$ bits. If $\rho=1-\epsilon$ with $\epsilon\le1/8$, then $\theta\ge1-4\epsilon$ and the bound is at least $(1-4\epsilon)\,n\log_2k-nH_2(4\epsilon)$.
\end{theorem}
\begin{proof}
We normalize all budgets to $1$. Let $w=\rho/2+1/4$, so $1/2<w<\rho$, $\rho-w=(2\rho-1)/4$ and $1-w=(3-2\rho)/4$. For $i=1,\dots,k$ let
\[
s_i=\tfrac12+(w-\tfrac12)\tfrac{k+1-i}{k+1},\qquad\text{so that}\qquad\tfrac12<s_k<\dots<s_1<w .
\]
For $t\in[k]$ let $J_t$ be the following sequence of queries for a single advertiser of budget $1$. It consists of the \emph{prefix} $s_1,\dots,s_t$ (as bids), then the \emph{follow-up} $f_t=1-s_t$, and then $k-t$ queries with bid $0$. The gadget has three properties. (a) Any two prefix bids sum to more than $1$, so at most one prefix item is accepted. (b) $\mathrm{OPT}(J_t)=1=s_t+f_t$. (c) If an algorithm accepts a prefix item $s_j$ with $j<t$, its remaining budget is $1-s_j<1-s_t=f_t$ (as $s_j>s_t$), so $f_t$ cannot be accepted and the revenue is $s_j<w$. If it accepts no prefix item, the revenue is at most $f_t<1/2<w$. Only if it accepts the last prefix item $s_t$ can the revenue exceed $w$, and it is at most $1$.

For $\mathbf t=(t_1,\dots,t_n)\in[k]^n$ let $I_{\mathbf t}$ present the gadgets $J_{t_1},\dots,J_{t_n}$ one after the other, gadget $i$ for advertiser $i$ with all its queries of degree one. All $I_{\mathbf t}$ have $m=n(k+1)$ queries and $\mathrm{OPT}=n$. We call block $i$ of $I_{\mathbf t}$ a \emph{hit} if the algorithm accepts the last prefix item of block $i$.

If $h$ blocks are hits then by (c) $\mathrm{ALG}\le h+(n-h)w$, with strict inequality if $h<n$. If $\mathcal A$ is $\rho$-competitive, then $\rho n\le\mathrm{ALG}$, hence $h(1-w)\ge n(\rho-w)$ and $h\ge\frac{\rho-w}{1-w}n=\theta n$, i.e., $h\ge c$.

Let $b$ be the maximum number of advice bits read by $\mathcal A$ on the family and $\sigma(\mathbf t)\in\{0,1\}^b$ the first $b$ bits of the tape used on $I_{\mathbf t}$. We fix $\sigma\in\{0,1\}^b$ and consider only those $\mathbf t$ with $\sigma(\mathbf t)=\sigma$. On these instances $\mathcal A$ behaves as with tape $\sigma0^\infty$. We fix $t_1,\dots,t_{i-1}$. By execution invariance the decisions of $\mathcal A$ on the prefix items of block $i$ do not depend on $t_i$, so the index $a_i\in[k]\cup\{\bot\}$ of the first prefix item accepted in block $i$ is determined by $\sigma$ and $t_1,\dots,t_{i-1}$ alone. Since at most one prefix item is accepted, block $i$ is a hit if and only if $t_i=a_i$. Consequently, for a fixed set $R\subseteq[n]$ of size $c$, at most $k^{n-c}$ vectors $\mathbf t$ with $\sigma(\mathbf t)=\sigma$ have all blocks in $R$ as hits, because the coordinates in $R$ are determined one after the other and the others are free. Every $\rho$-competitive instance has at least $c$ hits, so at most $\binom nck^{n-c}$ instances are served by $\sigma$. All $k^n$ instances must be served, hence $k^n\le2^b\binom nck^{n-c}$, i.e., $b\ge c\log_2k-\log_2\binom nc$. Finally $\binom nc\le2^n$ and $c\ge\theta n$. For $n=1$ we have $c=1$ and $\binom11=1$, so $2^b\ge k$. If $\rho=1-\epsilon$ with $\epsilon\le1/8$, then $\theta=\frac{1-2\epsilon}{1+2\epsilon}\ge1-4\epsilon$, so $c\ge(1-4\epsilon)n$ and $n-c\le4\epsilon n\le n/2$. Hence $\binom nc=\binom n{n-c}\le2^{nH_2(4\epsilon)}$, which gives the stated refinement. \qed
\end{proof}

\begin{theorem}
\label{thm:private}
Let $\epsilon\in(0,1/2]$ and let $m$ be known. For OAW-HB instances in which every query has degree one there is a deterministic online algorithm that is $(1-2\epsilon)$-competitive and uses at most
\[
n\Big[(\lceil1/\epsilon\rceil-1)\lceil\log_2m\rceil+2\lceil\log_2(\lceil1/\epsilon\rceil+1)\rceil\Big]=O\big(\tfrac n\epsilon\log m\big)
\]
bits of advice.
\end{theorem}
\begin{proof}
With degree-one queries the instance decomposes into $n$ independent single-advertiser instances, and $\mathrm{OPT}=\sum_i\mathrm{OPT}_i$, where $\mathrm{OPT}_i\le B_i$ is the largest subset sum of the usable bids of $i$ not exceeding $B_i$. We fix $i$ and an optimal subset $O_i$. Let $L_i=\{j\in O_i:b_{ij}>\epsilon B_i\}$, so that $|L_i|<1/\epsilon$, let $w_i=\sum_{j\in L_i}b_{ij}$, $\hat w_i=\lceil w_i/(\epsilon B_i)\rceil\in\{0,\dots,\lceil1/\epsilon\rceil\}$ and $\bar w_i=\hat w_i\epsilon B_i\in[w_i,w_i+\epsilon B_i)$. The advice for $i$ consists of $|L_i|$, the indices of the queries in $L_i$, and $\hat w_i$.

\emph{Algorithm for $i$.} The algorithm maintains the \emph{small spend} $\sigma_i$, initially $0$. On a query $j$ of $i$ it accepts if $j\in L_i$. Otherwise, if $b_{ij}\le\epsilon B_i$ and $\sigma_i+b_{ij}\le B_i-\bar w_i$, it accepts and sets $\sigma_i\leftarrow\sigma_i+b_{ij}$, and else it rejects.

At any time the spend on $i$ is at most $w_i+\sigma_i$, and $\sigma_i\le\max\{0,B_i-\bar w_i\}$. As $\bar w_i\ge w_i$, the spend never exceeds $B_i$, and every query in $L_i$ fits when it arrives.

We have $\mathrm{ALG}_i=w_i+\sigma_i^{\mathrm{fin}}$. If no query with $b_{ij}\le\epsilon B_i$ and $j\notin L_i$ was rejected, then all of $O_i\setminus L_i$ was accepted and $\mathrm{ALG}_i\ge\mathrm{OPT}_i$. Otherwise some such query $j_0$ was rejected because $\sigma_i+b_{ij_0}>B_i-\bar w_i$, whence $\sigma_i^{\mathrm{fin}}>B_i-\bar w_i-\epsilon B_i>B_i-w_i-2\epsilon B_i$ and $\mathrm{ALG}_i>(1-2\epsilon)B_i\ge(1-2\epsilon)\mathrm{OPT}_i$. Summing over $i$ proves the ratio. \qed
\end{proof}

\begin{theorem}
\label{thm:exact-large}
Let $s,\Delta,k\ge1$. There is a family of OAW-HB instances with $n=s\Delta$ advertisers of budget $2^{k+1}$, with $s(k+\Delta)$ queries of degree at most $\Delta$ and integer bids, on which every optimal deterministic algorithm reads at least $sk\log_2(\Delta+1)$ bits of advice on some instance. Conversely, on every instance with $m$ queries of degree at most $\Delta$, $\lceil m\log_2(\Delta+1)\rceil$ bits suffice for optimality.
\end{theorem}
\begin{proof}
The oracle fixes an optimal solution and writes, for every query, the position of its advertiser among the at most $\Delta$ eligible ones, or a rejection. These $m$ symbols form one number in base $\Delta+1$.

The advertisers form $s$ blocks of size $\Delta$, and $B=2^{k+1}$. In the first phase every block receives queries $j=1,\dots,k$, where query $j$ bids $2^{j-1}$ on each advertiser of the block. The first phase is the same in all instances. An instance is given by maps $f_r\colon[k]\to[\Delta]\cup\{\bot\}$, one for each block $r$. Let $s_{r,i}=\sum_{j\,:\,f_r(j)=i}2^{j-1}$. In the second phase advertiser $i$ of block $r$ receives one private query of bid $B-s_{r,i}$.

\emph{The optimum is unique in the first phase.} Assigning the first phase by the maps $f_r$, with $\bot$ meaning rejection, fills every advertiser exactly, so $\mathrm{OPT}=nB$, and a solution is optimal only if every advertiser ends exactly full. Fix an optimal solution and advertiser $i$ of block $r$, and let $\ell$ be its load after the first phase. The first phase of a block has total bid $2^k-1<B$, so the advertiser must accept its private query. This needs $\ell\le s_{r,i}$, and fullness then gives $\ell=s_{r,i}$. Since binary representations are unique, the set of first-phase queries assigned to $i$ is exactly $f_r^{-1}(i)$. Hence an optimal algorithm assigns the first phase exactly by the maps $f_r$.

An algorithm that reads at most $b$ bits is one of $2^b$ advice-free algorithms, and by execution invariance each of them has a single behaviour on the common first phase. The $(\Delta+1)^{sk}$ instances need pairwise different behaviours, so $2^b\ge(\Delta+1)^{sk}$. \qed
\end{proof}

\noindent For $s=1$ the family has $m=k+\Delta$ queries, so exact optimality costs $m\log_2(\Delta+1)$ bits up to an additive $\Delta\log_2(\Delta+1)+1$. The bids have $k$ bits, and with bids at most $W$ the theorem gives $\Omega(\frac n\Delta\log W\log(\Delta+1))$. For $\Delta=1$ and $s=1$ this is the known bound for online knapsack \cite{bockenhauer2012advice}.

\subsection{All Unit-Bid Instances}
\label{app:unit}

We consider instances in which every query has unit demand and bids $1$ on each advertiser of its eligible set, budgets are arbitrary integers, and queries are divisible. No structure of the eligible sets is assumed. For weights $\beta\in\mathbb R_{>0}^n$ the \emph{proportional allocation} sends the fraction $\beta_i/\sum_{i'\in N(j)}\beta_{i'}$ of query $j$ to advertiser $i\in N(j)$, and an advertiser accepts what it receives until it is full. This rule is online, and it earns $\sum_i\min\{B_i,A_i\}$, where $A_i$ is the total amount sent to $i$.

\begin{theorem}
\label{thm:unit}
Let $\epsilon\in(0,1]$. There is $R=O(\epsilon^{-2}\log\frac n\epsilon)$ such that a deterministic algorithm with $\lceil n\log_2(2R+1)\rceil$ bits of advice has ratio at least $1-\epsilon$ on every divisible unit-bid instance with $n$ advertisers. This is $2n\log_2\frac1\epsilon+n\log_2\log\frac n\epsilon+O(n)$ bits.
\end{theorem}
\begin{proof}
Agrawal, Zadimoghaddam and Mirrokni \cite[Thm.~1]{agrawal2018proportional} proved that for every instance there are weights of the form $\beta_i=(1+\epsilon')^{k_i}$ with integers $|k_i|\le R$ and $\epsilon'=\Theta(\epsilon)$, for which the proportional allocation earns at least $(1-\epsilon)$ times the optimal fractional value. These are the weights of their iterative procedure after $R$ rounds, which starts from $\beta_i=1$ and changes every weight by a factor $(1+\epsilon')^{\pm1}$ or not at all in each round. The optimal fractional value equals $\mathrm{OPT}$ because budgets are integers. The oracle writes $(k_1,\dots,k_n)$ as one number in base $2R+1$, and the algorithm runs the proportional allocation. \qed
\end{proof}

\noindent With Theorem~\ref{thm:nested-lb}, whose instances are of this type, the advice complexity of ratio $1-\epsilon$ on divisible unit-bid instances lies between $n\log_2\frac1\epsilon-O(n+\frac1\epsilon)$, for $\epsilon\le\frac14$, and $2n\log_2\frac1\epsilon+n\log_2\log\frac n\epsilon+O(n)$. So it is $\Theta(n\log\frac1\epsilon)$ whenever $\frac1n\le\epsilon\le\frac1{\log n}$, on all instances and not only on decreasing ones. For indivisible queries, assigning every query at random with the same probabilities loses a further factor $1-O(1/\sqrt{\min_iB_i})$ in expectation \cite[Remark~1]{agrawal2018proportional}. This uses large budgets, and for matching ($B_i=1$) the bounds $\Omega(n\log\frac1\epsilon)$ and $O(n/\epsilon^5)$ of \cite{durr2016power} remain the best known.
\end{document}